\documentclass[aps,pra,nofootinbib]{revtex4}
\usepackage{amsmath}
    \numberwithin{equation}{section}
\usepackage{graphicx}
\usepackage{dcolumn}
\usepackage{longtable}
\usepackage{array}
\usepackage{centernot}

\graphicspath{{figures/}}
\graphicspath{{diagrams/}}
\newcommand{\be}{\begin{equation}}
\newcommand{\ee}{\end{equation}}
\newcommand{\bearray}{\begin{eqnarray}}
\newcommand{\eearray}{\end{eqnarray}}
\newcommand{\bea}{\begin{eqnarray}}
\newcommand{\eea}{\end{eqnarray}}
\newcommand{\bse}{\begin{subequations}}
\newcommand{\ese}{\end{subequations}}
\newcommand{\bcases}{\begin{cases}}
\newcommand{\ecases}{\end{cases}}
\newcommand{\bpm}{\begin{pmatrix}}
\newcommand{\epm}{\end{pmatrix}}
\newcommand{\llangle}{\left\langle}
\newcommand{\rrangle}{\right\rangle}
\newcommand{\blangle}{\big\langle}
\newcommand{\brangle}{\big\rangle}

\newcommand{\hsm}{\hspace{-2pt}}

\newcommand{\vsq}[1]{\vec #1 \hsm^2}
\newcommand{\crr}{\nonumber \\}

\newcommand{\half}{\frac{1}{2}}

\def\dbar{{\mathchar'26\mkern-11mud}}

\begin{document}

\title{Coulomb expectation values in $D=3$ and $D=3-2\epsilon$ dimensions}

\author{Gregory S. Adkins}
\email[]{gadkins@fandm.edu}
\affiliation{Franklin \& Marshall College, Lancaster, Pennsylvania 17604}

\author{Md Faisal Alam}
\affiliation{Franklin \& Marshall College, Lancaster, Pennsylvania 17604}

\author{Conor Larison}
\affiliation{Franklin \& Marshall College, Lancaster, Pennsylvania 17604}

\author{Ruosi Sun}
\affiliation{Franklin \& Marshall College, Lancaster, Pennsylvania 17604}

\date{\today}

\begin{abstract}
We explore the quantum Coulomb problem for two-body bound states, in $D=3$ and $D=3-2\epsilon$ dimensions, in detail, and give an extensive list of expectation values that arise in the evaluation of QED corrections to bound state energies.  We describe the techniques used to obtain these expectation values and give general formulas for the evaluation of integrals involving associated Laguerre polynomials.  In addition, we give formulas for the evaluation of integrals involving subtracted associated Laguerre polynomials--those with low powers of the variable subtracted off--that arise when evaluating divergent expectation values.  We present perturbative results (in the parameter $\epsilon$) that show how bound state energies and wave functions in $D=3-2\epsilon$ dimensions differ from their $D=3$ dimensional counterparts and use these formulas to find regularized expressions for divergent expectation values such as $\blangle \bar V^3 \brangle$ and $\blangle (\bar V')^2 \brangle$ where $\bar V$ is the $D$-dimensional Coulomb potential.  We evaluate a number of finite $D$-dimensional expectation values such as $\blangle r^{-2+4\epsilon} \partial_r^2 \brangle$ and $\blangle r^{4\epsilon} p^4 \brangle$ that have $\epsilon \rightarrow 0$ limits that differ from their three-dimensional counterparts $\blangle r^{-2} \partial_r^2 \brangle$ and $\blangle p^4 \brangle$.  We explore the use of recursion relations, the Feynman-Hellmann theorem, and momentum space brackets combined with $D$-dimensional Fourier transformation for the evaluation of $D$-dimensional expectation values.  The results of this paper are useful when using dimensional regularization in the calculation of properties of Coulomb bound systems.
\end{abstract}

\maketitle


\section{Introduction}
\label{introduction}

The Coulomb force provides for the binding of all atoms.  The nature of the Coulomb force is most clearly displayed in the structure of two-body atoms such as hydrogen, hydrogen-like ions (He$^+$, Li$^{++}$, etc.), positronium ($e^+ e^-$), muonium ($\mu^+ e^-$), true-muonium ($\mu^+ \mu^-$), muonic hydrogen ($p^+ \mu^-$), and many more.  Because of its great importance, the quantum mechanics of the two-body Coulomb problem has been extensively studied.  In the course of the calculation of relativistic and QFT corrections to the energy levels of two-body Coulombic bound states, expectation values of various operators in quantum bound states are eventually required.  Many of these expectation values, for instance that of $1/r^3$, are singular due to short-distance divergences, and some regularization is required.  An extremely convenient scheme for obtaining finite quantities is provided by dimensional regularization.  In this work we discuss the dimensionally regularized Schr\"odinger equation for the Coulomb potential and show how the $D$-dimensional solution can be used to find regularized values for otherwise divergent quantities.  Another goal of this work is to present an extensive list of expectation values--both finite and divergent--that actually arise in practical calculations.  To obtain the finite ones, we developed a number of general integral formulas involving associated Laguerre polynomials, logarithms, and powers, and explored their consequences.  To obtain the divergent ones, we required a detailed understanding of the short-distance behavior of the $D$-dimensional wave functions.

This paper is organized as follows.  In Sec.~\ref{Coulomb_in_D} we discuss the quantum mechanical Coulomb problem in $D=3-2\epsilon$ dimensions, showing how to find the wave functions and energy levels and exploring their behavior near $\epsilon = 0$.  In Sec.~\ref{expectations_in_D_dims} we explore the general problem of finding expectation values in $D$ dimensions with a focus on divergent expectation values for which dimensional regularization is used.  We give an example of the evaluation of a divergent expectation value and show when the $\epsilon \rightarrow 0$ limit of a finite expectation value calculated in $D=3-2\epsilon$ dimensions might be different from the same finite expectation value calculated directly in three dimensions.  We also discuss some implications of the $D$-dimensional recursion relations and the Feynman-Hellmann theorem.  In Sec.~\ref{example} we illustrate the use of divergent Coulomb expectation values in a practical calculation.  In Sec.~\ref{Laguerre_integrals} we develop general formulas for the evaluation of integrals involving one or two associated Laguerre polynomials of the sort that arise in calculations involving Coulomb bound states.  Sec.~\ref{discussion} is a brief conclusion.  The main results of this paper, the expectation values themselves, appear in Appendix~\ref{expectation_values}, along with some of the closely related momentum space brackets.  The formula for $D$-dimensional Fourier transforms is given in Appendix~\ref{Fourier_transform}.  We have included a review of the three-dimensional quantum Coulomb problem in Appendix~\ref{Coulomb_in_3d}.  This has a focus on Coulomb bound states, and is included since the notation for associated Laguerre polynomials is not standardized, and because it is helpful to have a listing of useful results in one place and in a consistent notation.  Properties of the special functions used in this work are given in Appendix~\ref{special_functions}.  Useful formulas for gamma, polygamma, zeta, beta, and hypergeometric functions are included, as is a discussion of the harmonic and ``diharmonic'' numbers that appear in quantum Coulomb calculations.  Finally, we have given in Appendix~\ref{eval_of_bracket_lnq} a discussion of a specific momentum space bracket, $\blangle \! \ln q \brangle_{\vec p_2,\vec p_1}$, that is needed in calculations but is a special case.

Throughout this paper we use units where $\hbar=c=1$.  Standard units can be recovered at every stage by inserting the factors of $\hbar$ and $c$ needed to restore conventional dimensions.

The paper contains a mixture of new, known but obscure, and well-known results.  Results of all types are included with the goal of giving a unified and comprehensive presentation and useful tables.  Most of the $D=3-2\epsilon$ results are new although many of the momentum space brackets (see Sec.~ \ref{expectations_in_D_dims}) for $n=1$ were obtained in \cite{Czarnecki99}.  The application of the generalized power series solution for the $D=3-2\epsilon$ wave function was introduced in \cite{Adkins18} but is applied more systematically here.  Section \ref{Laguerre_integrals} involves integrals of associated Laguerre polynomials times powers of $x$ and $\ln x$.  Various expressions for the basic integral $\int_0^\infty dx \, e^{-x} x^s L_n^k(x) L_{n'}^{k'}(x)$ have appeared in the literature.  Our formula for this integral differs from others we have seen and is quite convenient for our purposes.  General formulas for the integrals involving logarithms are, to our knowledge, new, as are the formulas for integrals involving subtracted associated Laguerre polynomials.  We have not seen the summation formulas derived in Section \ref{Laguerre_integrals} elsewhere.  The definition and derived properties of the diharmonic sums given in Appendix \ref{special_functions} are, we believe, new.

\section{The quantum Coulomb problem in $D=3-2\epsilon$ dimensions}
\label{Coulomb_in_D}

In this section we review the quantum mechanics of the Coulomb problem in $D$ dimensions.  Dimensional regularization has been extremely useful for atomic physics calculations \cite{Brown92,Pineda98a,Pineda98b,Pineda98c,Czarnecki99a,Czarnecki99,Czarnecki05,Jentschura05,Wundt08,Adkins18,Adkins18a} and for quark-antiquark bound states involving heavy quarks (reviewed, for example in \cite{Hoang02,Brambilla05,Pineda12}).  The dimensionally regularized Coulomb problem serves as an extension of the usual three-dimensional case that allows us to evaluate quantities that would be divergent in three dimensions.  Central to our discussion is the Coulomb potential energy between an electron and a nucleus consisting of $Z$ positive charges.  In momentum space, this interaction energy is
\be \label{momentum_space_V}
\bar V(\vec k\,) = \frac{-4\pi Z \alpha \bar \mu^{2\epsilon}}{\vec k\,^2} ,
\ee
which can be obtained directly from the Coulomb-gauge Feynman rules for the interaction between the two charges with the exchange of a Coulomb photon.  The $\overline{MS}$ mass parameter that was introduced to keep $\alpha$ dimensionless can be written as
\be
{\overline{\mu}}^2 = \mu^2_{\overline{MS}} = \frac{\mu^2 e^{\gamma_E}}{4 \pi} ,
\ee
where $\mu$ is another parameter with units of mass and $\gamma_E$ is the Euler-Mascheroni constant.  The corresponding coordinate space potential energy is
\be
\bar V(r) = \int \dbar^D k \, e^{i \vec k \cdot \vec x} \bar V(\vec k\,) = - \frac{\Gamma \left (D/2-1 \right ) {\bar \mu}^{2 \epsilon} Z \alpha}{\pi^{D/2-1} r^{D-2}} = - \frac{\beta}{r^{1-2\epsilon}} ,
\ee
where $\dbar^n k$ is the $n$-dimensional momentum space integration measure $\dbar^n k = \frac{d^n k}{(2\pi)^n}$ and $\beta = \Gamma(D/2-1) \bar \mu^{2\epsilon} Z \alpha/\pi^{D/2-1}$.  (For the $D$-dimensional Fourier transform, see \cite{Samko78} and Appendix~\ref{Fourier_transform}.)  This potential energy satisfies the $D$-dimensional Gauss' Law
\be
-\vec \nabla\,^2 \bar V(r) = - 4 \pi Z \alpha \bar \mu^{2\epsilon} \delta^D(\vec x\,) .
\ee

The Schr\"odinger-Coulomb equation in $D$ dimensions is \cite{Andrew90,Morales96,Dong11,Bures15}
\be \label{Schroedinger_equation}
\frac{\vec p\,^2}{2 m_r} \psi(\vec x\,) + \bar V(r) \psi(\vec x\,) = \bar E \psi(\vec x\,)
\ee
where $m_r=m_1 m_2/(m_1+m_2)$ is the reduced mass.  Since the Coulomb potential has no angular dependence, it is possible to separate radial from angular variables as in three dimensions.  We write
\be
\psi(\vec x\,) = \bar R(r) Y(\hat x)
\ee
and use
\be
\vec \nabla^2 = \partial_r^2 + \frac{D-1}{r} \partial_r - \frac{L^2}{r^2}
\ee
where
\be
L^2 = \sum_{i<j} L_{i j}^2 \quad \text{with} \quad L_{i j} = -i \left ( x_i \partial_j - x_j \partial_i \right ) .
\ee
We choose $Y_\ell(\hat x)$ to be an eigenfunction of $L^2$ with eigenvalue $\ell(\ell+D-2)$ \cite{Avery89,Avery10}.  For example, the first few of these $D$-dimensional angular eigenfunctions are $Y_0(\hat x) \propto 1$, $Y_{1, i}(\hat x) \propto \hat x_i$, $Y_{2, i j}(\hat x) \propto \hat x_i \hat x_j - \frac{1}{D} \delta_{i j}$.
Consequently, the radial function satisfies
\be \label{radial_SCE_0}
\frac{1}{2 m_r} \left \{ - \partial_r^2 - \frac{D-1}{r} \partial_r + \frac{\ell (\ell+D-2)}{r^2} \right \} \bar R_{n \ell}(r) + \bar V(r) \bar R_{n \ell}(r) = \bar E_{n \ell} \bar R_{n \ell}(r) .
\ee
The quantum numbers listed for $\bar R$ include the angular momentum quantum number $\ell$ and a principal quantum number $n$ that is an integer $n=1$, 2, 3, etc. labeling the states with a given value of $\ell$ ordered by energy.  The energy $\bar E_{n \ell}$ for $D \ne 3$ depends on $\ell$ as well as $n$.

It is useful to factor out the leading short and long distance behavior of the wave function as in three dimensions.  The leading short distance terms are all part of $\vec \nabla^2$.  The $1/r^2$ term in $\vec \nabla^2$ dominates the $1/r^{D-2}$ of the potential and the constant in the energy term.  So, at short distances, $\bar R_{n \ell}$ satisfies
\be
\left \{ \partial_r^2 + \frac{D-1}{r} \partial_r - \frac{\ell (\ell+D-2)}{r^2} \right \} \bar R_{n \ell}(r) = 0 .
\ee
This differential equation is of Euler (or Cauchy-Euler) form and is satisfied by a power $\bar R_{n \ell}(r) = r^m$.  Insertion of this trial solution into the equation gives
\be
m(m-1) + (D-1) m - \ell(\ell+D-2) = (m-\ell)(m+\ell+D-2) = 0 ,
\ee
so $m=\ell$ or $m=-(\ell+D-2)$.  For $\ell \ge 0$ and $D$ near three the second solution is negative and consequently unacceptable.  The implied behavior of the radial function near $r=0$ is
\be
\bar R_{n \ell}(r) \rightarrow r^\ell .
\ee
For long distances the leading terms in (\ref{radial_SCE_0}) are the first and the last:
\be
\frac{1}{2 m_r} \left \{ - \partial_r^2 \right \} \bar R_{n \ell}(r) = E_{n \ell} \bar R_{n \ell}(r) ;
\ee
the solution that falls to zero at large distances is
\be
\bar R_{n \ell}(r) \rightarrow e^{-\bar \gamma_{n \ell} r} 
\ee
where
\be \label{eqn_energy_from_gamma}
\bar E_{n \ell} = -\frac{\bar \gamma_{n \ell}^2}{2 m_r} , \quad \bar \gamma_{n \ell} = \sqrt{-2 m_r \bar E_{n \ell}} \; .
\ee
We define a new function $L_{n \ell}(\rho)$ according to
\be
\bar R_{n \ell}(r) = \bar \phi_{n \ell} \Omega_{D-1}^{1/2} \left ( \frac{(n+\ell)!}{n (n-\ell-1)!} \right )^{1/2} \frac{\rho^\ell}{(2\ell+1)!} e^{-\rho/2} L_{n \ell}(\rho)
\ee
where $2 \bar \gamma_{n \ell} r \equiv \rho$, $\bar \phi_{n \ell}$ is a normalization constant such that
\be
\bar \phi_n  \equiv \bar \phi_{n 0} = \lim_{r \rightarrow 0} \bar \psi_{n 0}(r) 
\ee
is the S-state wave function at the origin (at ``contact''), $\bar \phi_{n \ell}$ with $\ell>0$ is related to the $\ell^\text{th}$ derivative of the wave function at contact, and
\be
\Omega_{n} = \frac{2 \pi^{\frac{n+1}{2}}}{\Gamma \left ( \frac{n+1}{2} \right )}
\ee
is the surface area of a unit $n$-sphere.  That is, $\Omega_{D-1}$ is the total solid angle of $D$-dimensional space.   The new function $L_{n \ell}(\rho)$ is the analog of the associated Laguerre polynomial $L_{n-\ell-1}^{2\ell+1}(\rho)$ but is normalized differently:
\be
\lim_{\rho \rightarrow 0} L_{n \ell}(\rho) = 1 .
\ee
In the three-dimensional limit one has
\be
\lim_{\epsilon \rightarrow 0} L_{n \ell}(\rho) = {\binom{n+\ell}{2\ell+1}}^{-1} L_{n-\ell-1}^{2\ell+1}(\rho) \quad \text{and} \quad \lim_{\epsilon \rightarrow 0} \bar \phi_{n \ell} = \left ( \frac{(m_r Z \alpha)^3}{\pi n^3} \right )^{1/2} .
\ee
When expressed in terms of $L_{n \ell}(\rho)$, the radial equation (\ref{radial_SCE_0}) becomes
\be \label{radial_SCE_1}
\left \{ \partial_r^2 + \left ( \frac{2 \ell+D-1}{r} - 2 \bar \gamma \right ) \partial_r - \frac{(2 \ell +D -1) \bar \gamma}{r} -2 m_r \bar V(r) \right \} L_{n \ell}(\rho) = 0 .
\ee
It seems natural to change the independent variable from $r$ to the dimensionless variable $\rho$.  In doing so we will be dividing equation (\ref{radial_SCE_1}) by $(2 \bar \gamma)^2$.  The potential term becomes
\be
\frac{-2 m_r}{(2 \bar \gamma_{n \ell})^2} \bar V(r) = \frac{2 m_r \Gamma(1/2-\epsilon) {\bar \mu}^{2\epsilon} Z \alpha}{(2 \bar \gamma_{n \ell}) \pi^{1/2-\epsilon} (2 \bar \gamma_{n \ell} r) } \left ( \frac{\rho}{2 \bar \gamma_{n \ell}} \right )^{2 \epsilon} = \frac{\bar{n}_{n \ell} \rho^{2 \epsilon}}{\rho} ,
\ee
where
\be \label{eqn_for_nbar_1}
\bar{n}_{n \ell} \equiv \frac{m_r Z \alpha}{\bar \gamma_{n \ell}} \frac{\Gamma(1/2-\epsilon)}{\pi^{1/2-\epsilon}} \left ( \frac{\bar \mu}{2 \bar \gamma_{n \ell}} \right )^{2 \epsilon} ,
\ee
which would be an integer $n$ with $n \ge 1$ in three dimensions.  The radial equation with a dimensionless independent variable becomes
\be \label{radial_SCE_2}
\left \{ \partial_\rho^2 + \left ( \frac{2 (\ell+1-\epsilon)}{\rho} - 1 \right ) \partial_\rho - \frac{\ell+1-\epsilon}{\rho} + \frac{\bar{n}_{n \ell} \rho^{2 \epsilon}}{\rho} \right \} L_{n \ell}(\rho) = 0 .
\ee
A bit of trial and error reveals that a power series expansion around the origin is more complicated than usual and takes the form
\bea
L_{n \ell}(\rho) &=& 1 + \sum_{j=1}^\infty a_{j 0} \rho^j + \sum_{j=1}^\infty a_{j 1} \bar{n}_{n \ell} \rho^{j+2 \epsilon} + \sum_{j=2}^\infty a_{j 2} \bar{n}_{n \ell}^2 \rho^{j+4 \epsilon} + \cdots \crr
&=& \sum_{k=0}^\infty \sum_{j=k}^\infty a_{j k} \bar{n}_{n \ell}^k \rho^{j+2 \epsilon k}
= \sum_{j=0}^\infty \sum_{k=0}^j a_{j k} \bar{n}_{n \ell}^k \rho^{j+2 \epsilon k}
\eea
with $a_{0 0}=1$ and $a_{j k}=0$ unless $0 \le k \le j$.  We insert $L_{n \ell}(\rho)$ into (\ref{radial_SCE_2}) and assume that all powers $\rho^{j+2 \epsilon k}$ are independent.  We isolate the coefficient of $\rho^{j-2+2 \epsilon k}$, which is required to vanish, and obtain the following recursion relation
\be
a_{j k}  = \frac{a_{j-1, k} (j+\ell+\epsilon (2k-1)) - a_{j-1, k-1} }{(j+2 \epsilon k) (j+2 \ell+1+2 \epsilon (k-1))} .
\ee
The recursion relation, along with the boundary conditions listed above, allows us to find all of the coefficients $a_{j k}$.  For example, we find that $a_{1 0} = \frac{1}{2}$, $a_{1 1} = \frac{-1}{2(1+\ell)(1+2\epsilon)}$, etc.  It would be easy to write a routine to calculate as many of these coefficients as desired.

As a quick consistency check we consider the three-dimensional limit and calculate the total coefficient $A_j$ of a given power $\rho^j$.  This coefficient is
\be
A_j = \sum_{k=0}^j a_{j k} n^k = \frac{(-1)^j \Pi_{k=1}^j ( n-\ell-k )}{j! \Pi_{k=1}^j (2\ell+1+k)} = \frac{(-1)^j (n-\ell-1)! (2\ell+1)!}{j! (n-\ell-j-1)! (2\ell+1+j)!} ,
\ee
where we have used $\epsilon \rightarrow 0$, $\bar{n}_{n \ell} \rightarrow n$.  We see that when $n=\ell+k$ where $k$ is a positive integer, the series terminates with a maximum power of $\rho^{n-\ell-1}$.  By comparison with (\ref{eqn_assoc_Laguerre_series}), we see that $L_{n \ell}$ in this limit is just proportional to the usual associated Laguerre polynomials.

The differential equation (\ref{radial_SCE_2}) can be solved numerically (as in \cite{Adkins18}) to find the wave function and eigenvalues $\bar n_{n \ell}$ (which give values for $\bar  \gamma_{n \ell}$ using (\ref{eqn_for_nbar_1}) and then for the energies using (\ref{eqn_energy_from_gamma})), or a perturbative approach can be employed.  The perturbative method employs a $D$-dimensional lowest-order problem with potential $\tilde V(r)=-Z \alpha/r$, for which an exact solution is known \cite{Alliluev58,Nieto79}.  Building a perturbation scheme on the basis of this lowest order problem, with 
\be
H' = \bar V(r)-\tilde V(r) = - \frac{2 Z \alpha}{r} \big ( \ln(\mu r) + \gamma_E \big ) \epsilon + O(\epsilon^2) ,
\ee
one can obtain expressions for the energies
\be
\bar E_{n \ell} = E_n \left \{ 1 + \epsilon \left [ 4 \ln \left ( \frac{\mu n}{2 m_r Z \alpha} \right ) + 4 \text{H}_{n+\ell} + \frac{2}{n} \right ] + O(\epsilon^2) \right \} ,
\ee
where $E_n = - \frac{m_r (Z \alpha)^2}{2 n^2}$ represents the usual Bohr energies, and for the ``wave function at the origin" factor
\bea
\bar \phi_{n \ell} &=& \Omega_{D-1}^{-1/2} \left ( \frac{n n_r!}{(n+\ell)!} \right )^{1/2} \frac{(2\ell+1)!}{(2 \bar \gamma_{n \ell})^\ell} \lim_{r \rightarrow 0} \frac{1}{r^\ell} \bar R_{n \ell}(r) \crr
&=& \left ( \frac{\gamma_n^D}{\pi} \right )^{1/2} \bigg \{ 1 + \epsilon \bigg [ 3 \ln \left ( \frac{\mu n}{2 m_r Z \alpha} \right ) + 2 n \, \text{diH}_+(n+\ell,-n_r) - n \left ( \text{H}^2_{n+\ell} - \text{H}^{(2)}_{n+\ell} \right ) \crr
&\hbox{}& \hspace{0.0cm} + n \left ( \text{H}_{n_r}^2 + \text{H}_{n_r}^{(2)} \right ) + 2 \text{H}_{n+\ell} + 2 \text{H}_{2\ell+1} + \frac{1}{2} \left ( \ln \pi - \gamma_E \right ) - 2 + \frac{2}{n} - 2n \zeta(2) \bigg ] + O(\epsilon^2) \bigg \} ,
\eea
where $n_r = n-\ell-1$ is the radial quantum number, $\gamma_n \equiv \frac{m_r Z \alpha}{n}$, and the harmonic, generalized harmonic, and diharmonic numbers $\text{H}_n$, $\text{H}^{(2)}_n$, and $\text{diH}_\pm(n,m)$ are defined in Appendix \ref{special_functions}.

\section{Coulomb expectation values in $D=3$ and $D=3-2\epsilon$ dimensions}
\label{expectations_in_D_dims}

Expectation values in $D$ dimensions are defined in the usual way:
\be
\blangle M \brangle = \int d^D x \, \psi^\dagger(\vec x\,) M(\vec x\,) \psi(\vec x\,) ,
\ee
where $M$ is an operator that may involve derivatives as well as functions of position.  Many of the expectation values we will need are finite in three dimensions, but others, such as $\blangle \bar V^3 \brangle \propto \blangle r^{-3+6\epsilon} \brangle$ and $\blangle (\bar V')^2 \brangle \propto \blangle r^{-4+4\epsilon} \brangle$, are divergent for S states.  These divergences are controlled by dimensional regularization, and working in $D$ dimensions we can find their explicit values.  In this section we will give some details of the evaluation of Coulomb expectation values--with a specific focus on the evaluation of divergent ones.  Situations can also arise for finite expectation values where taking the $D \rightarrow 3$ limit and taking the expectation value do not commute:
\be
\llangle \lim_{D \rightarrow 3} M \rrangle \ne \lim_{D \rightarrow 3} \blangle M \brangle .
\ee
We will give an example of this subtlety and describe when it will occur.  Some special techniques are useful for finding and/or testing expectation values, such as recursion relations and the Feynman-Hellmann theorem.  We will explore some consequences of these methods in three and $D$ dimensions.  Finally, we will discuss momentum space brackets and their relation to expectation values.

Many innovative procedures for the evaluation of Coulomb expectation values have appeared in the literature, including those of \cite{Schroedinger26,Waller26,VanVleck34,Pasternack37,Bethe57,Epstein62,Srivastava03,Suslov08,Gonzalez17}, as well as useful lists of expectation values: \cite{Swainson91a,Titard94,Czarnecki99,Zatorski08}.  Our lists overlap with these but are more extensive.

The expectation value $\blangle (\bar V')^2 \brangle_{n \ell}$ is divergent for $\ell=0$ with a divergence proportional to $\blangle r^{-4} \brangle_{n 0}$ in three dimensions, so it must be regularized in order to be evaluated.  When $\ell>0$ this expectation value is finite due to the extra factors of $r^\ell$ in the wave function, and we can use standard three-dimensional methods as described in Sec.~\ref{Laguerre_integrals}.  After performing the angular integrals, the S state expectation value can be written as
\be
\blangle (\bar V')^2 \brangle_{n 0} = \int_0^\infty dr \, r^{D-1} \bar R_{n 0}(r) \big [ \bar V'(r) \big ]^2 \bar R_{n 0}(r)
\ee
where $\bar R_{n 0}(r) = \bar \phi_{n 0} \Omega_{D-1}^{1/2} e^{-\rho/2} L_{n 0}(\rho)$ with $2 \bar \gamma_{n \ell} r = \rho$.  The $L_{n 0}(\rho)$ function is a generalized series
\be
L_{n 0}(\rho) = 1 + \frac{1}{2} \rho - \frac{\bar n_{n 0}}{2(1+2\epsilon)} \rho^{1+2\epsilon} + O(\rho^2) .
\ee
Only the first three terms in $L_{n 0}(\rho)$--the ones shown--are sensitive at short distances, so we isolate the short distance part $\hat L_{n 0} \equiv 1+\rho/2-\bar n_{n 0} \rho^{1+2\epsilon}/(2[1+2\epsilon])$ for separate treatment.  The remaining terms in the series, starting at $O(\rho^2)$, we define as ${^2\!}L_{n 0}(\rho) = L_{n 0}(\rho)-\hat L_{n 0}(\rho)$.  We write the expectation value with $D \rightarrow 3-2\epsilon$ as
\be
\blangle (\bar V')^2 \brangle_{n 0} = \bar \phi_{n 0}^2 \Omega_{D-1} \int_0^\infty dr \, r^{2-2\epsilon} e^{-\rho} \left [ (1-2\epsilon) \beta r^{-2+2\epsilon} \right ]^2 \left \{ \big ( \hat L_{n 0} \big )^2 + \hat L_{n 0} \, {^2\!}L_{n 0} + {^2\!}L_{n 0} L_{n 0} \right \}
\ee
where $\bar V(r) = -\beta/r^{1-2\epsilon}$.  Only the first term involving $\big ( \hat L_{n 0} \big )^2$ is divergent (when $\epsilon=0$), but for $\text{Re}(\epsilon)>1/2$ its $r$ integral is convergent and can be expressed in terms of gamma functions.  We analytically continue the result to a neighborhood of $\epsilon=0$ and expand in a Laurent series to obtain a series with a $1/\epsilon$ pole expressing the divergence plus a series with non-negative powers of $\epsilon$.  The remaining terms containing ${^2\!}L_{n 0}$ are finite and can be evaluated with $\epsilon \rightarrow 0$ using the methods of Sec.~\ref{Laguerre_integrals}.  In all we find
\be
\llangle (\bar V')^2 \rrangle_{n 0} = \pi m_r (Z \alpha)^3 \bar \phi_{n 0}^2 \bar \mu^{2\epsilon} \left \{ -\frac{2}{\epsilon} - 8 \ln \left ( \frac{\mu n}{2 m_r Z \alpha} \right ) + 8 \text{H}_n + \frac{4}{3n^2} - \frac{4}{n} - \frac{16}{3} + O(\epsilon) \right \} .
\ee
The expectation value $\blangle \bar V^3 \brangle_{n 0}$ is simpler because only the first term in the series for $L_{n 0}$ must be included in $\hat L_{n 0}$.  For other expectation values as many terms should be included in $\hat L_{n 0}$ as needed.

Extra care is required for some finite expectation values for which the operations of taking the expectation value and the $D \rightarrow 3$ limit do not commute.  An example is $\blangle r^{-2+4\epsilon} \partial_r^2 \brangle_{n 0}$.  When taking the second derivative of the radial part of the wave function, second derivatives of $L_{n 0}$ occur, which look like
\be
\partial_\rho^2 L_{n 0}(\rho) = - \bar n_{n 0}\epsilon \rho^{-1+2\epsilon} + O(\rho^0) .
\ee
Negative powers of $\rho$ generated in this way can lead to divergences in the radial integral that would not have been present in three dimensions where the associated Laguerre is a regular polynomial.  The radial integral gives a $1/\epsilon$ divergence, which along with the explicit factor of $\epsilon$ produced by the differentiation, leads to an additive constant that would not have appeared in three dimensions.  The same separation of $L_{n 0}$ into $\hat L_{n 0}+{^2\!}L_{n 0}$ allows us to evaluate this expectation value safely, finding
\be
\blangle r^{-2+4\epsilon} \partial_r^2 \brangle_{n \ell} = (m_r Z \alpha)^4 \left \{ \frac{-2n^2-1+2\ell(\ell+1)}{4(\ell-1/2)(\ell+1/2)(\ell+3/2)n^5} - \frac{2}{n^3} \delta_{\ell=0} \right \} ,
\ee
which differs from $\blangle r^{-2} \partial_r^2 \brangle_{n 0}$ evaluated directly in three dimensions (as tabulated in (\ref{expec_rm2dr2})) only in the $\delta_{\ell=0}$ term.

Expectation values involving the momentum squared positioned next to a wave function (or adjoint wave function) may be simplified by use of the Schr\"odinger equation.  We write (\ref{Schroedinger_equation}) in the form $p^2 \psi = 2 m_r (H-\bar V) \psi$, so that
\bea
\blangle M p^2 \brangle &=& 2 m_r \blangle M ( E- \bar V ) \brangle , \\[3pt]
\blangle p^2 M \brangle &=& 2 m_r \blangle (E- \bar V) M \brangle 
\eea
for any operator $M$.

Recursion relations (\cite{Pasternack37,Epstein62}) allow us to express one expectation value in terms of others, which may already be known.  Recursion relations can be found by noting that for any operator $A$, the expectation value of $B \equiv [H,A]$ vanishes: $\blangle B \brangle=0$.  When we deal with radial operators only, we can use the radial Hamiltonian
\be \label{eqn_radial_Hamiltonian}
H = \frac{1}{2m_r} \left ( - \partial_r^2 - \frac{D-1}{r} \partial_r + \frac{\ell(\ell+D-2)}{r^2} \right ) - \frac{\beta}{r^{1-2\epsilon}} .
\ee  
Using $A \rightarrow r^s$ we find $0 = s(s+1-2\epsilon) \blangle r^{s-2} \brangle + 2 s \blangle r^{s-1} \partial_r \brangle$,
or
\be \label{eqn_recursion_for_rsdr}
\blangle r^s \partial_r \brangle = - \frac{1}{2} ( s+2-2\epsilon) \blangle r^{s-1} \brangle \quad \text{when} \; s \ne -1 .
\ee
Now taking $A \rightarrow r^s \partial_r$ and using (\ref{eqn_radial_Hamiltonian}) to write $\partial_r^2$ in terms of $H$ and eliminating $\blangle r^{s-2} \partial_r \brangle$ using (\ref{eqn_recursion_for_rsdr}), we find
\be \label{eqn_recursion_for_rs}
0 = 8 m_r \bar E (s+1) \blangle r^s \brangle + 4 m_r \beta (2s+1+2\epsilon) \blangle r^{s-1+2\epsilon} \brangle + s \big [ s^2-(1-2\epsilon)^2-4\ell(\ell+1-2\epsilon) \big ] \blangle r^{s-2} \brangle .
\ee
Additional relations could be found by making other choices for $A$.  We can now derive some consequences.  Use of (\ref{eqn_recursion_for_rsdr}) with $s=-2+2\epsilon$ gives
\be
\blangle \bar V' \partial_r \brangle = 0 
\ee
(except when $\ell=0$, since $\blangle r^{-D} \brangle$ is a special case not regulated by dimensional regularization as noted below (\ref{eqn_FT_of_lnq}); the $\ell=0$ result for $\blangle \bar V' \partial_r \brangle$ can be obtained using the Fundamental Theorem of Calculus and is given in (\ref{eqn_expec_Vbarp_dr})).  Use of (\ref{eqn_recursion_for_rsdr}) with  $s=-3+4\epsilon$ gives
\be
\blangle \bar V \bar V' \partial_r \brangle = -\frac{1}{2} \blangle (\bar V')^2 \brangle .
\ee
Use of (\ref{eqn_recursion_for_rs}) with $s=0$ and $\langle 1 \rangle = 1$ gives
\be \label{eqn_Vbar_and_Ebar}
\blangle \bar V \brangle = \frac{2 \bar E}{1+2\epsilon} .
\ee
Use of (\ref{eqn_recursion_for_rs}) with $s=-2+4\epsilon$ gives
\be \label{eqn_recursion_for_m2p4eps}
4 m_r \bar E (1-4\epsilon) \blangle \bar V^2 \brangle - 2 m_r (3-10\epsilon) \blangle \bar V^3 \brangle + \left ( 3 (1-2\epsilon) - \frac{4\ell(\ell+1-2\epsilon)}{1-2\epsilon} \right ) \blangle (\bar V')^2 \brangle = 0 .
\ee
These results are exact in $D$ dimensions.

The Feynman-Hellmann theorem \cite[p.195]{Weinberg15} gives additional relationships involving expectation values.  For $\lambda$ a parameter of the Hamiltonian, the theorem states that
\be
\llangle \frac{\partial H}{\partial \lambda} \rrangle = \frac{\partial \bar E}{\partial \lambda} .
\ee
Using the radial Hamiltonian of (\ref{eqn_radial_Hamiltonian}), the relevant parameters are $m_r$, $\beta$, and $\ell$.  The equation for $m_r$ is
\be \label{eqn_FH_mr}
\llangle - \frac{H-\bar V}{m_r} \rrangle = \frac{\partial \bar E}{\partial m_r} ,
\ee
and the equation for $\beta$ is
\be \label{eqn_FH_beta}
\llangle \frac{\bar V}{\beta} \rrangle = \frac{\partial \bar E}{\partial \beta} .
\ee
The first order differential equations implied by (\ref{eqn_Vbar_and_Ebar}), (\ref{eqn_FH_mr}) and (\ref{eqn_FH_beta}) lead to 
\be
\bar E \propto m_r^{\frac{1-2\epsilon}{1+2\epsilon}} \beta^\frac{2}{1+2\epsilon} ,
\ee
which can also be seen using (\ref{eqn_energy_from_gamma}) and (\ref{eqn_for_nbar_1}).  The equation for $\ell$ leads to a value for $\blangle 1/r^2 \brangle$ in three dimensions, but its utility in $D$ dimensions is unclear to us.

Momentum space brackets \cite{Czarnecki99} are defined according to
\be
\Bigl \langle M(\vec p_2,\vec p_1) \Bigr \rangle_{\vec p_2,\vec p_1} \equiv \int \dbar^D p_2 \, \dbar^D p_1 \, \psi^\dagger (\vec p_2\,) M(\vec p_2,\vec p_1\,) \psi(\vec p_1\,) 
\ee
where $\dbar^D p = d^D p/(2\pi)^D$.  Momentum space brackets are useful because Feynman rules are often expressed in momentum space and such brackets emerge naturally in the initial stages of a calculation.  Fourier transformation can be employed to give an equivalent coordinate space version
\bearray
\left \langle M(\vec p_2,\vec p_1) \right \rangle_{\vec p_2,\vec p_1} &=&
\int d^D x_2 \, d^D x_1 \, \dbar^D p_2 \, \dbar^D p_1 \, \psi^\dagger (\vec x_2) e^{i \vec p_2 \cdot \vec x_2} M(\vec p_2,\vec p_1) e^{-i \vec p_1 \cdot \vec x_1} \psi(\vec x_1) \crr
&=& \int d^D x_2 \, d^D x_1 \, \psi^\dagger(\vec x_2) M(\vec x_2,\vec x_1) \psi(\vec x_1) ,
\eearray
where $M(\vec x_2,\vec x_1)$ is the Fourier transform 
\be 
M(\vec x_2,\vec x_1) \equiv \int \dbar^D p_2 \, \dbar^D p_1 \, e^{i \vec p_2 \cdot \vec x_2} M(\vec p_2,\vec p_1) e^{-i \vec p_1 \cdot \vec x_1} .
\ee
When $M(\vec p_2,\vec p_1)$ can be expressed entirely in terms of the relative momentum $\vec q=\vec p_2-\vec p_1$, one can write $M(\vec p_2,\vec p_1)=M(\vec q\,)$ and $M(\vec x_2,\vec x_1)$ takes the form
\be
M(\vec x_2,\vec x_1) \equiv \int \dbar^D q \, \dbar^D p_1 \, e^{i \vec q \cdot \vec x_2} M(\vec q\,) e^{i \vec p_1 \cdot (\vec x_2 - \vec x_1)} 
= M(\vec x_2) \delta^D(\vec x_2-\vec x_1) .
\ee
In that case, the momentum space bracket is equal to a corresponding expectation value
\be
\Bigl \langle M(\vec p_2,\vec p_1) \Bigr \rangle_{\vec p_2,\vec p_1} = \int d^D x \, \psi^\dagger(\vec x\,) M(\vec x\,) \psi(\vec x\,) = \left \langle M \right \rangle 
\ee
where $M(\vec x\,)=\int \dbar^D q \, e^{i \vec q \cdot \vec x} M(\vec q\,)$ (see Appendix~\ref{Fourier_transform}) is the Fourier transform of $M(\vec p_2,\vec p_1)=M(\vec q\,)$.
Brackets involving components of $\vec p_2$ or $\vec p_1$ in the numerator can be dealt with by replacing $p_{1 i}$, for example, by $i \partial_{1 i}$ acting on the exponential $e^{-i \vec p_1 \cdot \vec x_1}$ followed by a partial integration to let the derivative act on the wave function $\psi(\vec x_1)$.  One finds, for instance, that the momentum space bracket of $p_{2 i} N(\vec q\,) p_{1 i}$ can be written as an expectation value according to
\be
\blangle p_{2 i} N(\vec q\,) p_{1 i} \brangle_{\vec p_2,\vec p_1} = \blangle p_i N(\vec x\,) p_i \brangle ,
\ee
where $p_i=-i \partial_i$ and $N(\vec x\,)$ is the Fourier transform of $N(\vec q\,)$.

The momentum space Schr\"odinger equation
\be
\frac{\vec p\,^2}{2m_r} \psi(\vec p\,) + \int \dbar^D k \, \bar V(\vec p-\vec k\,) \psi(\vec k\,) = \bar E \psi(\vec p\,)
\ee
where $\bar V(\vec p-\vec k\,)=-4 \pi Z \alpha \bar \mu^{2\epsilon}/(\vec p-\vec k\,)^2$ can be extremely useful for simplifying momentum space brackets directly in momentum space.  We find that
\be
\Bigl \langle M(\vec p_2,\vec p_1) \vsq{p_1} \Bigr \rangle_{\vec p_2,\vec p_1} = 2 m_r \int \dbar^D p_2 \, \dbar^D p_1 \, \psi^\dagger (\vec p_2\,) \bigg \{ M(\vec p_2,\vec p_1\,) \bar E - \int \dbar^D k \, M(\vec p_2,\vec k\,) \bar V(\vec k-\vec p_1) \bigg \} \psi(\vec p_1\,) ,
\ee
with an extra integral to do that may lead to an overall simplification.  An analogous relation holds for $\blangle \vsq{p_2} M(\vec p_2,\vec p_1) \brangle_{\vec p_2,\vec p_1}$.

A number of expectation values and momentum space brackets are presented in Appendix~\ref{expectation_values}.  Many of the three-dimensional ones appear elsewhere, but we thought it would be useful to have a complete list containing both finite and divergent expectation values in one place.  The finite ones, where possible, have been checked using numerical integration in three dimensions for many values of the quantum numbers, and we have verified that all of them are consistent with the various recursion relations and other identities.

A few of the expectation values in the lists are singular when evaluated directly in three dimensions but are finite in the $D \rightarrow 3$ limit of the $D$-dimensional result.  One of these is $\blangle p^6 \brangle$.  After use of the Schr\"odinger equation on the left and right, we find
\be
\blangle p^6 \brangle = (2m_r)^2 \big \{ \bar E^2 \blangle \vec p\,^2 \brangle - \bar E \blangle \vec p\,^2 \bar V \brangle - \bar E \blangle \bar V \vec p\,^2 \brangle + \blangle \bar V \vec p\,^2 \bar V \brangle \big \} .
\ee
The final expectation value here is divergent in three dimensions as can best be seen in momentum space:
\be
\blangle \bar V \vec p\,^2 \bar V \brangle = \llangle \int \dbar^D k \, \bar V(\vec p_2-\vec k\,) \vec k\,^2 \bar V(\vec k-\vec p_1) \rrangle_{\vec p_2,\vec p_1}
= \llangle \int \dbar^D k \, \frac{(-4 \pi Z \alpha \bar \mu^{2\epsilon})}{(\vec p_2-\vec k\,)^2} \vec k\,^2 \frac{(-4 \pi Z \alpha \mu^{2\epsilon})}{(\vec k-\vec p_1)^2} \rrangle_{\vec p_2,\vec p_1} ,
\ee
where the $\vec k$ integral is divergent for large $\vert \vec k\, \vert$ in three dimensions.  In $D$ dimensions, for appropriate values of $D$, the $\vec k$ integral is well-defined and can be performed by use of a Feynman parameter.  We find that
\be
\blangle \bar V \vec p\,^2 \bar V \brangle = 2^{1+2\epsilon} \pi^{3/2-\epsilon} \beta^2 \frac{\Gamma(1/2+\epsilon)}{\Gamma(1-2\epsilon)} \llangle \frac{\vec p_2 \cdot \vec p_1 }{(\vec q\,^2)^{1/2+\epsilon}} \rrangle_{\vec p_2,\vec p_1} ,
\ee
which has a finite $D \rightarrow 3$ limit.  We write this in coordinate space using the Fourier transform formula of Appendix~\ref{Fourier_transform} as
\be
\blangle \bar V \vec p\,^2 \bar V \brangle = \blangle p_i \bar V^2 p_i \brangle ,
\ee
and note that the three dimensional expectation value $\blangle p_i V^2 p_i \brangle$ has the finite value (\ref{expec_V_p2_V}) that can be obtained from (\ref{p_rminus2_p}).  The final regularized value of $\blangle p^6 \brangle$ is given in (\ref{expec_p6}).  Another exceptional case is the momentum space bracket $\blangle 1/\vec q\,^4 \brangle_{\vec p_2,\vec p_1}$, which is divergent in three dimensions, but when evaluated in $D$ dimensions has a finite $D \rightarrow 3$ limit.  We use the $D$-dimensional Fourier transform of Appendix~\ref{Fourier_transform} to see that
\be \llangle \frac{1}{\vec q\,^4} \rrangle_{\vec p_2,\vec p_1} = \frac{\Gamma(-1/2-\epsilon)}{16 \pi^{D/2}} \blangle r^{1+2\epsilon} \brangle \xrightarrow[\epsilon \to 0]{} - \frac{1}{8 \pi} \blangle r \brangle = - \frac{1}{16\pi} \Bigl \{ 3n^2-\ell(\ell+1) \Bigr \} (m_r Z \alpha)^{-1} .
\ee

One particular bracket bears special mention--that of $\ln q$ where $q = \vert \vec q \, \vert$.  This bracket appears in calculations, but its evaluation has particular difficulties.  The bracket can be expressed as an expectation value using the Fourier transform ($\text{FT}$) of $\ln q$
\be
\blangle \! \ln q \brangle_{\vec p_2,\vec p_1} = \blangle \text{FT}[\ln q](\vec x\,) \brangle .
\ee
We write $\ln q$ as $\ln q = \displaystyle{\lim_{s \rightarrow 0}} \frac{d}{ds} q^s$.  If we assume that the operation of taking the limit of a derivative commutes with the operation of taking a Fourier transform (done using the formulas of Appendix~\ref{Fourier_transform}), we find that the transform of $\ln q$ is 
\be \label{eqn_FT_of_lnq}
\text{FT}[\ln q](\vec x\,) \rightarrow \lim_{s \rightarrow 0} \frac{d}{ds} \text{FT}[q^s](\vec x\,) = - \frac{\Gamma \left ( \frac{D}{2} \right )}{2 \pi^{D/2} r^D} .
\ee
This function $1/r^D$ is problematic because it gives divergent S state expectation values that are not regulated by dimensional regularization: $\int_0^\infty dr \, r^{D-1} \bar R_{n 0}(r) r^{-D} \bar R_{n 0}(r)$ has a short distance divergence.  It seems that our usual approach of evaluating momentum space brackets by Fourier transformation followed by calculating a coordinate space expectation value must fail.  We could fall back on momentum space calculations, which work for individual states, but it seems difficult to find a general formula (for all states) by using momentum space methods.  Instead, we evaluate the bracket of $\ln q$ by keeping $s$ as the regulating parameter and hold off on taking the $s$ derivative and $s \rightarrow 0$ limit until after the expectation value is done.  This procedure gives a finite result for $\blangle \! \ln q \brangle_{\vec p_2,\vec p_1}$ that agrees with direct evaluations of momentum space brackets for particular states.  Details of the calculation are given in Appendix~\ref{eval_of_bracket_lnq}.

\section{Evaluation of an example of an energy level contribution}
\label{example}

\begin{figure}
\includegraphics[width=3.5in]{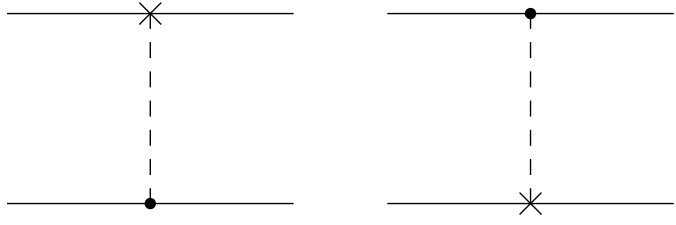}
\caption{\label{coulomb_CX1} Diagrams showing the $CX1$ corrections to bound state energies.  These graphs contain a Coulomb photon exchange (dashed line) between electron and nucleus (solid lines).  The $CX1$ interaction is shown as a cross, while the usual Coulomb interaction ($-i q$) is shown as a dot.  The bound state wave functions on the left and right are present but not represented in the diagram.}
\end{figure}

We give a sample calculation of a dimensionally regularized contribution to Coulomb energy levels as an illustration of the usefulness of Coulomb expectation values of the sort studied and tabulated in this work.  We base our calculation on the dimensionally regularized low-energy effective quantum field theory Non-Relativistic QED (NRQED) \cite{Caswell86}.  We consider the contribution illustrated in Fig.~\ref{coulomb_CX1}, which shows a process involving exchange of a Coulomb photon having a Coulomb vertex on one end and a higher-order vertex on the other.  In particular, we will calculate the effect of the ``$CX1$'' higher-order vertex, which comes from the term
\be
\delta {\cal L} = \psi^\dagger c_{X1} \frac{q}{m^4} \big [ \vec D\,^2 , \vec D \cdot \vec E + \vec E \cdot \vec D \big ] \psi
\ee
in the NRQED Lagrangian (as given in \cite{Hill13}).  The factor $c_{X1}$ is a Wilson coefficient ({\it i.e.} a matching coefficient) and lends its name to this contribution.  The fermion fields $\psi^\dagger$ and $\psi$ are two-component Pauli spinors representing the electron or the ``nucleus'' (assumed also to be a spin-1/2 fermion) with charge $q$ and mass $m$.  The photon field is contained in the electric field $\vec E=-\vec \nabla A^0 - \partial_0 \vec A$, and $\vec D = \vec \nabla - i q \vec A$ is the  gauge covariant derivative.  The Feynman rule from $\delta {\cal L}$ for the interaction of a fermion with a Coulomb photon is $c_{X1} \frac{-i q}{m^4} \left ( \vsq{p_2} - \vsq{p_1} \right )^2$.  A formalism for using NRQED to calculate bound state energy levels is given in \cite{Adkins18a}.  It makes use of the NRQED Bethe-Salpeter equation and a perturbative scheme based on a lowest-order problem that can be reduced to the usual Schr\"odinger equation with a ($D=3-2\epsilon$ dimensional) Coulomb potential.  The Bethe-Salpeter wave functions in this formalism are $\bar \Psi(p)$ and $\Psi(p)$, and the energy shift coming from the two diagrams of Fig.~\ref{coulomb_CX1} is
\be \label{energy1}
\Delta E_{CX1} = i \int \dbar^d p_2 \, \dbar^d p_1 \, \bar \Psi(p_2) \left \{ c^{(1)}_{X1} \frac{-i q_1}{m_1^2} \left ( \vsq{p_2}-\vsq{p_1} \right )^2 (-i q_2) + (-i q_1) c^{(2)}_{X2} \frac{-i q_2}{m_1^2} \left ( \vsq{p_2}-\vsq{p_1} \right )^2 \right \} \frac{i}{(\vec p_2 - \vec p_1\,)^2} \Psi(p_1) ,
\ee
where $d=4-2\epsilon$ is the dimension of spacetime.  The $c_{X1}$ matching coefficient might be different for different particles.  For elementary spin 1/2 particles such as the electron and muon (and their antiparticles) it is $c_{X1}=\frac{5}{128}+O(\alpha)$ \cite{Hill13}.

The reduction of (\ref{energy1}) to an expectation value and then to a number happens through several steps.  We perform the energy integrals by use of a convenient property of the lowest-order NRQED Bethe-Salpeter wave functions: $\int \dbar p_{0} \, \bar \Psi(p) = \psi^\dagger(\vec p\,)$ and $\int \dbar p_0 \Psi(p) = \psi(\vec p\,)$, where $\psi(\vec p\,)$ is the spatial wave function discussed in Sec.~\ref{Coulomb_in_D}.  The $CX1$ energy correction simplifies to
\bea
\Delta E_{CX1} &=& -4 \pi Z \alpha \bar \mu^{2\epsilon} \left ( \frac{c_{X1}^{(1)}}{m_1^4} + \frac{c_{X2}^{(2)}}{m_2} \right ) \int \dbar^D p_2 \, \dbar^D p_1 \, \psi^\dagger(\vec p_2) \frac{(\vsq{p_2} - \vsq{p_1})^2}{\vec q\,^2} \psi(\vec p_1) 
\eea
where the integrals now are over the $D=3-2\epsilon$ spatial dimensions, the charge product is $q_1 q_2 = - 4\pi Z \alpha \bar \mu^{2\epsilon}$ as in (\ref{momentum_space_V}), and we define $\vec q \equiv \vec p_2-\vec p_1$.  We expand the numerator factor $(\vsq{p_2}-\vsq{p_1})^2 = \vec p_2 \hsm^4 - 2 \vsq{p_2} \vsq{p_1} + \vec p_1 \hsm^4$.  We use the Fourier transform as discussed in Sec.~\ref{expectations_in_D_dims} and Appendix~\ref{Fourier_transform} to turn our expression for the energy correction into a standard (D-dimensional) expectation value:
\be
\Delta E_{CX1} = \left ( \frac{c_{X1}^{(1)}}{m_1^4} + \frac{c_{X2}^{(2)}}{m_2} \right ) \blangle \vec p\,^4 \bar V - 2 \vec p\,^2 \bar V \vec p\,^2 + \bar V \vec p\,^4 \brangle .
\ee
The required expectation values have been tabulated as (\ref{expec_p4barV}) and (\ref{expec_p2barVp2}).  Or, we could use the identity of (\ref{identity_Vbarp2}) to write $\blangle \vec p\,^4 \bar V - 2 \vec p\,^2 \bar V \vec p\,^2 + \bar V \vec p\,^4 \brangle$ as $-4 m_r \blangle (\bar V')^2 \brangle$, and so reduce our result for the $CX1$ energy correction to
\bea
\Delta E_{CX1} &=& - 4 m_r \left ( \frac{c_{X1}^{(1)}}{m_1^4} + \frac{c_{X2}^{(2)}}{m_2} \right ) \blangle (\bar V' )^2 \brangle \crr
&\bbox{}& \hspace{-1.6cm} = - 4 m_r \left ( \frac{c_{X1}^{(1)}}{m_1^4} + \frac{c_{X2}^{(2)}}{m_2} \right ) \bcases \pi \bar \phi_n^2 m_r (Z \alpha)^3 \bar \mu^{2 \epsilon}
\left \{ - \frac{2}{\epsilon} - 8 \ln \left ( \frac{\mu n}{2 m_r Z \alpha} \right ) + 8 \text{H}_n + \frac{4}{3 n^2} - \frac{4}{n} - \frac{16}{3} \right \} & \text{if } \ell = 0
 \\[3pt]
\frac{3n^2-\ell(\ell+1)}{2 \ell (\ell+1) (\ell-1/2) (\ell+1/2) (\ell+3/2) n^5} m_r^4 (Z \alpha)^6 & \text{if } \ell > 0 \ecases ,
\eea
where we have used the value for $\blangle (\bar V')^2 \brangle$ tabulated in (\ref{expec_Vp2}).  It is extremely convenient to have results for needed expectation values readily available.

\section{Integrals and matrix elements involving associated Laguerre polynomials}
\label{Laguerre_integrals}

Many expectation values in $D=3-2\epsilon$ dimensions are finite and reduce to their three-dimensional form in the limit $\epsilon \rightarrow 0$.  In this section we develop the techniques for evaluating the types of integrals involving standard three-dimensional associated Laguerre polynomials that occur in the quantum mechanical Coulomb problem.  Where possible, general formulas are obtained, and a number of special cases most useful for atomic physics calculations are discussed in detail.  Some useful sums that follow from the integration formulas are given as well.

We begin by finding general expressions for the integrals 
\bse \bea
I_s(n,k) &=& \int_0^\infty dx \, e^{-x} x^s L_n^k(x) , \\[3pt]
J_s(n,k) &=& \int_0^\infty dx \, e^{-x} x^s \ln x L_n^k(x) , \\[3pt]
K_s(n,k; n',k') &=& \int_0^\infty dx \, e^{-x} x^s L_n^k(x) L_{n'}^{k'}(x) , \\[3pt]
L_s(n,k;n',k') &=& \int_0^\infty dx \, e^{-x} x^s \ln x \; L_n^k(x) L_{n'}^{k'}(x) , \\[3pt]
M_s(n,k;n',k') &=& \int_0^\infty dx \, e^{-x} x^s \ln^2 x \; L_n^k(x) L_{n'}^{k'}(x) ,
\eea \ese
(where $s$ must satisfy $s > -1$ in order to avoid a divergence for small $x$) and some related sums.  It is also occasionally necessary to find values for integrals like those above but with negative values of $s$ for which the integrals are divergent.  In that case subtractions are required--we choose to modify one of the associated Laguerre polynomials in the integrand by removing its first $p$ powers of $x$ where $p+s > -1$.  If we write
\be \label{expansion_for_LaguerreLnkx}
L_n^k(x) = \sum_{r=0}^n c_{n k r} x^r 
\ee
for the associated Laguerre function where $c_{n k r} = \frac{(-1)^r}{r!} \binom{n+k}{n-r}$ as in (\ref{eqn_assoc_Laguerre_series}), then we can define
\be
{}^p \! L_n^k(x) = L_n^k(x) - \sum_{r=0}^{p-1} c_{n k r} x^r = \sum_{r=p}^n c_{n k r} x^r .
\ee
The subtracted integrals are convergent for $p+s > -1$
\bse \bea
{}^p \! I_s(n,k) &=& \int_0^\infty dx \, e^{-x} x^s \; {}^p \! L_n^k(x) , \\[3pt]
{}^p \! J_s(n,k) &=& \int_0^\infty dx \, e^{-x} x^s \ln x \; {}^p \! L_n^k(x) , \\[3pt]
{}^p \! K_s(n,k; n',k') &=& \int_0^\infty dx \, e^{-x} x^s \; {}^p \! L_n^k(x) L_{n'}^{k'}(x) , \\[3pt]
{}^p \! L_s(n,k;n',k') &=& \int_0^\infty dx \, e^{-x} x^s \ln x \; {}^p \! L_n^k(x) L_{n'}^{k'}(x) , \\[3pt]
{}^p \! M_s(n,k;n',k') &=& \int_0^\infty dx \, e^{-x} x^s \ln^2 x \; {}^p \! L_n^k(x) L_{n'}^{k'}(x) .
\eea \ese

We can immediately obtain a result for $I_s(n,k)$ using the series expansion for $L_n^k(x)$ as
\bea \label{integral_Isnk_0}
I_s(n,k) &=& \int_0^\infty dx \, e^{-x} x^s L_n^k(x) \crr
&=& \sum_{r=0}^n c_{n k r} \int_0^\infty dx \, e^{-x} x^{s+r} \crr
&=& \sum_{r=0}^n \frac{(-1)^r \Gamma(n+k+1)}{(n-r)! \Gamma(k+r+1) r!} \Gamma(s+r+1) .
\eea
However, a more convenient form without a sum is possible through use of integration by parts.  We use the Rodrigues formula \eqref{Rodrigues} for $L_n^k(x)$ and define $D \equiv d/dx$ to write
\be
I_s(n,k) = \frac{1}{n!} \int_0^\infty dx \, x^{s-k} D^n \left ( x^{n+k} e^{-x} \right ) .
\ee
Now we integrate by parts $n$ times using the assumption that the real part of $s$ is sufficiently large that all integrated terms vanish at $x=0$.  They vanish for $x \rightarrow \infty$ due to the exponential factor.  One finds that
\bea \label{integral_Isnk}
I_s(n,k) &=& \frac{1}{n!} \int_0^\infty dx \, (-1)^n D^n \left ( x^{s-k} \right ) x^{n+k} e^{-x} \crr
&=& \frac{(-1)^n}{n!} \int_0^\infty dx \, \frac{\Gamma(s-k+1)}{\Gamma(s-k-n+1)} x^{s-k-n} x^{n+k} e^{-x} \crr
&=& \frac{(-1)^n}{n!} \frac{\Gamma(s-k+1) \Gamma(s+1)}{\Gamma(s-k-n+1)} .
\eea
Results for $I_s(n,k)$ with smaller values of $s$ are obtained by analytical continuation and possibly the appropriate limiting case as $s$ approaches the desired value.  For example, the integral $I_0(n-1,1)$ with $n \ge 1$ has the value
\bea \label{integral_I0nm11}
I_0(n-1,1) &=& \int_0^\infty dx \, e^{-x} L_{n-1}^1(x) \crr
&=& \lim_{s \rightarrow 0} \frac{(-1)^{n-1}}{(n-1)!} \frac{\Gamma(s) \Gamma(s+1)}{\Gamma(s-(n-1))} \crr
&=& 1 ,
\eea
where (\ref{gamma_fraction_identity_c}) was used to see that
\be \label{gamma_fraction_identity_c_1}
\lim_{s \rightarrow 0} \frac{\Gamma(s)}{\Gamma(s-N)} = (-1)^N N! .
\ee
Other integrals, similarly obtained, are
\be \label{integral_I1nm11}
I_1(n-1,1) = \int_0^\infty dx \, e^{-x} x L_{n-1}^1(x) = \frac{(-1)^{n-1}}{(n-1)!} \lim_{s \rightarrow 1} \frac{\Gamma(s) \Gamma(s+1)}{\Gamma(s-(n-1))} = \delta_{n=1} 
\ee
and
\be \label{integral_I1nm22}
I_1(n-2,2) = \int_0^\infty dx \, e^{-x} x L_{n-2}^2(x) = \frac{(-1)^{n-2}}{(n-2)!} \lim_{s \rightarrow 1} \frac{\Gamma(s-1) \Gamma(s)}{\Gamma((s-1)-(n-2))} = \delta_{n\ge 2}. 
\ee
A general result is
\be
I_s(n,k) = \begin{cases} \frac{s!(n+k-s-1)!}{n! (k-s-1)!} & 0 \le s \le k-1 \\[3pt] 0 & k \le s \le n+k-1 \\[3pt]
\frac{(-1)^n (s-k)! s!}{n! (s-n-k)!} & n+k \le s \end{cases} ,
\ee
where we assume that $s$, $k$, and $n$ are all non-negative integers.

We can obtain results for the subtracted integrals ${}^p \! I_s(n,k)$ starting from (\ref{integral_Isnk_0}).  We find
\be
{}^p \! I_s(n,k) = \sum_{r=p}^n \frac{(-1)^r \Gamma(n+k+1)}{(n-r)! \Gamma(k+r+1) r!} \Gamma(s+r+1) .
\ee
The sum can be done, for example by use of Mathematica \cite{Mathematica11.1}, giving
\be
{}^p \! I_s(n,k) = \frac{(-1)^p (n+k)! \Gamma(p+s+1)}{(n-p)! p! \Gamma(k+p+1)} {_3 F_2}(1,-n+p,1+p+s; 1+p,1+k+p;1) .
\ee
(The hypergeometric functions are defined and some properties are given in Appendix~\ref{special_functions}.)  For specific values of $p$ (with $p \geq 0$) this can be simplified.  As examples, one has
\bse \bea
{}^1 \! I_s(n,k) &=& \frac{\Gamma(n+k+1) \Gamma(s+1)}{n! \Gamma(k+1)} \Big \{ {_2 F_1}(-n,s+1;k+1;1)-1 \Big \} , \\[3pt]
{}^2 \! I_s(n,k) &=& \frac{\Gamma(n+k+1) \Gamma(s+1)}{n! \Gamma(k+1)} \Big \{ {_2 F_1}(-n,s+1;k+1;1)-1 - \frac{n(s+1)}{k+1} \Big \} ,
\eea \ese
where values having $s$ an integer in the range $-p \leq s \leq -1$ must be interpreted as limits.  We use (\ref{hypergeometric_identity_1}) to rewrite these as
\bse \bea
{}^1 \! I_s(n,k) &=& \frac{\Gamma(n+k+1) \Gamma(s+1)}{n! \Gamma(k+1)} \left \{ \frac{\Gamma(n+k-s) \Gamma(k+1)}{\Gamma(k-s) \Gamma(n+k+1)} - 1 \right \} , \\[3pt]
{}^2 I_s(n,k) &=& \frac{\Gamma(n+k+1) \Gamma(s+1)}{n! \Gamma(k+1)} \left \{ \frac{\Gamma(n+k-s) \Gamma(k+1)}{\Gamma(k-s) \Gamma(n+k+1)} - 1 + \frac{n(s+1)}{k+1} \right \} ,
\eea \ese
interpreting the formulas as limits as appropriate.
For the once subtracted integral with integral $s$, $s \ge -p=-1$ and $n \ge 1$, one has
\be \label{int_Itildesnm111}
{}^1 \! I_s(n-1,1) = \begin{cases} n \left ( 1 - \text{H}_n \right ) & s=-1 \\
1-n & s=0 \\
- s! \, n & 1 \le s \le n-1 \\
- s!  \, n + (-1)^{n-1} s! \binom{s-1}{n-1} & n \le s \end{cases} 
\ee
and with $n \ge 2$
\be
{}^1 \! I_s(n-2,2) = \begin{cases} \frac{1}{4} n (n-1) (3-2\text{H}_n) & s=-1 \\
- \frac{1}{2} (n-1)(n-2) & s=0 \\
-\frac{1}{2} (n+1)(n-2) & s=1 \\
-s! \frac{1}{2} n (n-1) & 2 \le s \le n-1 \\
-s! \frac{1}{2} n(n-1) + (-1)^n s! \binom{s-2}{n-2} & n \le s  \end{cases} .
\ee
For the twice subtracted integral with integral $s$, $s \ge -p=-2$, one has
\be \label{int_Itildesnm112}
{}^2 \! I_s(n-1,1) = \begin{cases} \frac{n(n+1)}{2} \left ( \frac{3}{n+1} + \text{H}_n - \frac{5}{2} \right ) & s=-2 \\
n \left ( \frac{n+1}{2}-\text{H}_n \right ) & s=-1 \\
 \frac{(n-1)(n-2)}{2} & s=0 \\
  \frac{s! n}{2} \left [ (s+1)n-(s+3) \right ] & 1 \le s \le n-1 \\
\frac{s! n}{2} \left [ (s+1)n-(s+3) \right ] + (-1)^{n-1} s! \binom{s-1}{n-1} & n \le s \end{cases} ,
\ee
all of which vanish automatically when $n=1$ or $n=2$ since $^2 \! L_{n-1}^1(x)$ vanishes in those cases.

The $J_s(n,k)$ integrals can be obtained from the $I_s(n,k)$ ones using
\be
\frac{d}{ds} x^s = \frac{d}{ds}e^{s \ln x} = x^s \ln x 
\ee
and
\be
\frac{d}{ds} \frac{\Gamma(s+a) \Gamma(s+b)}{\Gamma(s+c)} = \frac{\Gamma(s+a) \Gamma(s+b)}{\Gamma(s+c)} \Bigl \{ \psi(s+a) + \psi(s+b) - \psi(s+c) \Bigr \} .
\ee
One finds immediately that
\bea
J_s(n,k) &=& \int_0^\infty dx \, e^{-x} x^s \ln x L_n^k(x) \crr
&=& \frac{(-1)^n}{n!} \frac{\Gamma(s-k+1) \Gamma(s+1)}{\Gamma(s-k-n+1)} \Bigl \{ \psi( s-k+1) + \psi(s+1) - \psi(s-k-n+1) \Bigr \} ,
\eea
where as in $I_s(n,k)$ a limiting procedure for $s$ is often required.  The result for $J_s(n,k)$ can be simplified somewhat using the digamma identity (\ref{digamma_falling_identity}):
\be
J_s(n,k) = \frac{(-1)^n}{n!} \frac{\Gamma(s-k+1) \Gamma(s+1)}{\Gamma(s-k-n+1)} \left \{ \psi(s+1) - \sum_{r=1}^n \frac{1}{k-1+r-s} \right \} .
\ee
As an example we calculate $J_0(n-1,1) = \lim_{s \rightarrow 0} J_s(n-1,1)$:
\be \label{integral_J0nm11}
\int_0^\infty dx \, e^{-x} \ln x \, L_{n-1}^1(x) = \frac{(-1)^{n-1}}{(n-1)!} \lim_{s \rightarrow 0} \frac{\Gamma(s) \Gamma(s+1)}{\Gamma(s-(n-1))} \left \{ \psi(s+1) - \sum_{r=1}^{n-1} \frac{1}{r-s} \right \} = -\gamma_E - \text{H}_{n-1} ,
\ee
and more generally one has
\be
J_s(n-1,1) = \frac{(-1)^s}{(n-1)!} (s-1)! s! (n-1-s)! \quad \text{for $n \ge 2$ and $s$ an integer $1 \le s \le n-1$}.
\ee
Some of the $J_s(n-3,5)$ integrals are also useful, and one has
\be
J_s(n-3,5) = \frac{(n+1-s)! s!}{(n-3)! (4-s)!} \Bigl \{ \text{H}_s -\gamma_E - \text{H}_{n+1-s} + \text{H}_{4-s} \Bigr \}  \quad \text{for $n \ge 3$ and $s$ an integer $0 \le s \le 4$}.
\ee
A more general result is
\be
J_s(n,k) = \begin{cases} \frac{s!(n+k-s-1)!}{n! (k-s-1)!} \Bigl \{ \text{H}_s + \text{H}_{k-s-1} - \text{H}_{n+k-s-1} - \gamma_E \Bigr \} & 0 \le s \le k-1 \\[3pt] \frac{(-1)^{s-k+1} (s-k)! s! (n+k-s-1)!}{n!} & k \le s \le n+k-1 \\[3pt]
\frac{(-1)^n (s-k)! s!}{n! (s-n-k)!} \Bigl \{ \text{H}_s + \text{H}_{s-k} - \text{H}_{s-n-k} - \gamma_E \Bigr \} & n+k \le s \end{cases} ,
\ee
where we assume that $s$, $k$, and $n$ are all non-negative integers.

A comparison of (\ref{integral_Isnk_0}) with (\ref{integral_Isnk}) and a similar comparison for the logarithmic integrals results in a number of useful summation formulas.  One finds that
\be
\sum_{j=0}^n \frac{(-1)^j (n+k)!}{(n-j)! (k+j)! j!} \Gamma(s+j+1) = \frac{(-1)^n}{n!} \frac{\Gamma(s-k+1) \Gamma(s+1)}{\Gamma(s-k-n+1)} 
\ee
for non-negative integral values of $n$ and $k$ and any $s$, perhaps with a limit required.  For example, with $n \rightarrow n-1$, $k \rightarrow 1$, and $s \rightarrow 0$, one has
\be
\sum_{j=0}^{n-1} (-1)^j \binom{n}{j+1} = 1 .
\ee
For the logarithmic case, we return to the evaluation of $J_s(n,k)$ by the direct method:
\bea \label{integral_Jsnk_0}
J_s(n,k) &=& \int_0^\infty dx \, e^{-x} x^s \ln x \, L_n^k(x) \crr
&=& \sum_{j=0}^n c_{n k j} \int_0^\infty dx \, e^{-x} \ln x \, x^{s+j} \crr
&=& \sum_{j=0}^n \frac{(-1)^j (n+k)!}{(n-j)! (k+j)! j!} \Gamma(s+j+1) \psi(s+j+1) .
\eea
So one has the summation formula
\be \label{sum_formula_2}
\sum_{j=0}^n \frac{(-1)^j (n+k)!}{(n-j)! (k+j)! j!} \Gamma(s+j+1) \psi(s+j+1) 
= \frac{(-1)^n}{n!} \frac{\Gamma(s-k+1) \Gamma(s+1)}{\Gamma(s-k-n+1)} \left \{ \psi(s+1) - \sum_{r=1}^n \frac{1}{k-1+r-s} \right \} ,
\ee
valid for non-negative integral values of $n$ and $k$, and $s > -1$, perhaps with a limit required.  For example, with $n \rightarrow n-1$, $k \rightarrow 1$, $s \rightarrow 0$, one has
\be \label{summation_formula_for_MPBB_1}
\sum_{j=0}^{n-1} (-1)^j \binom{n}{j+1} \psi(j+1) = \psi(1) - \text{H}_{n-1} = - \gamma_E - \text{H}_{n-1} .
\ee
Also, with $n \ge 1$, $k \rightarrow 0$, and $s \rightarrow 0$, one finds
\be
\sum_{j=0}^n (-1)^j \binom{n}{j} \psi(j+1) = \frac{(-1)^n}{n!} \lim_{s \rightarrow 0} \frac{\Gamma(s+1) \Gamma(s+1)}{\Gamma(s-(n-1))} \left \{ \psi(s+1) - \sum_{r=1}^n \frac{1}{r-1-s} \right \} .
\ee
The denominator gamma factor is singular for all such $n$, so the whole term would vanish except if there is a pole in the curly bracket.  All terms there are finite except when $r=1$, in which case they go as $1/s$, and (\ref{gamma_fraction_identity_c_1}) leads to
\be
\sum_{j=0}^n (-1)^j \binom{n}{j} \psi(j+1) = -\frac{1}{n} .
\ee

For the evaluation of $K_s(n,k;n',k')$ we use the expansion (\ref{expansion_for_LaguerreLnkx}) for one of the associated Laguerre functions and the Rodrigues formula \eqref{Rodrigues} for the second and find
\bearray \label{gen_Laguerre_int} 
K_s(n,k; n',k') &=& \sum_{r=0}^n \frac{c_{n k r}}{n'!} \int_0^\infty dx \, e^{-x} x^s x^r e^x x^{-k'} D^{n'} \left ( x^{n'+k'} e^{-x} \right ) \cr
&=& \sum_{r=0}^n \frac{c_{n k r}}{n'!} \int_0^\infty dx \, (-1)^{n'} D^{n'} \left ( x^{s+r-k'} \right ) x^{n'+k'} e^{-x} \cr
&=& \sum_{r=0}^n \frac{c_{n k r}}{n'!} (-1)^{n'} \int_0^\infty dx \, \frac{\Gamma(s+r-k'+1)}{\Gamma(s+r-k'-n'+1)} x^{s+r-k'-n'} x^{n'+k'} e^{-x} \cr
&=& \frac{(n+k)! (-1)^{n'}}{n'!} \sum_{r=0}^n \frac{(-1)^r \Gamma(s+r+1-k') \Gamma(s+r+1)}{r! (n-r)! (k+r)! \Gamma(s+r+1-k'-n')} .
\eearray
The integration by parts above was done under the assumption that $Re(s)$ is large enough so that the integrated terms vanish for $x \rightarrow 0$ as in the derivation of (\ref{integral_Isnk}).  Integrals for smaller values of $s$ are obtained through analytic continuation and, if necessary, a limiting procedure for $s$.

The subtracted $K$ integrals differ from the above only by having a different lower limit for the $r$ summation:
\be \label{Laguerre_Ktilde_int}
{}^p \! K_s(n,k; n',k') = \frac{(n+k)! (-1)^{n'}}{n'!} \sum_{r=p}^n \frac{(-1)^r \Gamma(s+r+1-k') \Gamma(s+r+1)}{r! (n-r)! (k+r)! \Gamma(s+r+1-k'-n')} ,
\ee
and converge whenever $s+p > -1$.

We can use (\ref{gen_Laguerre_int}) to prove the Laguerre orthogonality relation (\ref{Laguerre_orthogonality}):
\bearray 
K_k(n,k;m,k) &=& \frac{(n+k)! (-1)^m}{m!} \sum_{r=0}^n \frac{(-1)^r \Gamma(r+1) \Gamma(k+r+1)}{r! (n-r)! (k+r)! \Gamma(r+1-m)} \cr
&=& \frac{(n+k)! (-1)^m}{m!} \sum_{r=0}^n \frac{(-1)^r}{(n-r)! \Gamma(r+1-m)} . 
\eearray
The gamma function in the denominator is infinite for every r in the sum when $m>n$, so $K_k(n,k;m,k)$ vanishes in this case.  By symmetry under $n \leftrightarrow m$, $K_k(n,k;m,k)$ also vanishes for $m<n$.  When $m=n$, the only nonvanishing contribution comes from $r=m=n$, with the result
\be
K_k(n,k;m,k) = \frac{(n+k)!}{n!} \delta_{n m} . 
\ee
The orthogonality relation (\ref{Laguerre_orthogonality})
\be 
\int_0^\infty dx \, e^{-x} x^k L_n^k(x) L_m^k(x) =  K_k(n,k;m,k) = \frac{(n+k)!}{n!} \delta_{n m} . 
\ee
follows immediately.

Another integral useful for normalizing the Coulomb wave functions is
\be \int_0^\infty dx \, e^{-x} x^{k+1} \left [ L_n^k(x) \right ]^2 = K_{k+1}(n,k;n,k) 
= \frac{(n+k)! (-1)^n}{n!} \sum_{r=0}^n \frac{(-1)^r \Gamma(r+2) \Gamma(k+r+2)}{r! (n-r)! \Gamma(k+r+1) \Gamma(r+2-n)} . \ee
Because of the function $\Gamma(r+2-n)$ in the denominator, the only nonvanishing terms are those with $r=n-1$ and $r=n$.  One finds that
\bearray \int_0^\infty dx \, e^{-x} x^{k+1} \left [ L_n^k(x) \right ]^2
&=& \frac{(n+k)! (-1)^n}{n!} \left \{ \frac{(-1)^n (n+1)! (k+n+1)!}{n! (k+n)!} + \frac{(-1)^{n-1} n! (k+n)!}{(n-1)! (k+n-1)!} \right \} \crr
&=& \frac{(n+k)!}{n!} (2n+k+1) .
\eearray
Another set of useful $K$ integrals is $K_1(n-1,1;n-a,a)$ with $a$ an integer $1 \le a \le n$, for which we use (\ref{gen_Laguerre_int}):
\be
K_1(n-1,1;n-a,a) = \frac{n! (-1)^{n-a}}{(n-a)!} \lim_{s \rightarrow 0} \sum_{r=0}^{n-1} \frac{(-1)^r \Gamma(s+r+1-a)}{r! (n-1-r)! \Gamma(s-(n-r-1))}
\ee
For $0 \le r \le a-1$ there is a divergent gamma function in both the numerator and the denominator (with a finite limit), while for $a \le r \le n-1$ there is a divergent gamma function only in the denominator.  It follows that
\bea
K_1(n-1,1;n-a,a) &=& \frac{n! (-1)^{n-a}}{(n-a)!} \sum_{r=0}^{a-1} \frac{(-1)^r} {r! (n-1-r)!}  \lim_{s \rightarrow 0} \frac{ \Gamma(s-(a-r-1))}{\Gamma(s-(n-r-1))} \crr
&=& \frac{n! (-1)^{n-a}}{(n-a)!} \sum_{r=0}^{a-1} \frac{(-1)^r} {r! (n-r-1)!} \frac{(-1)^{n-r-1} (n-r-1)!}{(-1)^{a-r-1} (a-r-1)!} \crr
&=& \frac{n!}{(n-a)! (a-1)!} \sum_{r=0}^{a-1} (-1)^r \binom{a-1}{r} \crr
&=& n \, \delta_{a=1} .
\eea
A similar calculation leads to
\be
K_0(n-1,1;n-a,a) = \binom{n}{a} \big \{ 1 - \delta_{a=0} \big \} .
\ee
We also find
\be
K_0(n-a,a+1;n-b,b+1) = \frac{(n+1)!}{(n-b)!} \sum_{r=0}^{b} \frac{(-1)^r (n-r)!}{(n-a-r)! (r+a+1)! (b-r)!} ,
\ee
and
\be
K_1(n-a,a+1;n-b,b+1) = \frac{(n+1)!}{(n-b)!} \sum_{r=0}^{b-1} \frac{(-1)^r (r+1) (n-1-r)!}{(n-a-r)! (r+a+1)! (b-1-r)!} .
\ee
As a summary of results for the $K$ integrals one has
\bse \bea
\label{integral_Kknkmk} K_k(n,k;m,k) &=& \frac{(n+k)!}{n!} \delta_{n m} , \\[3pt]
K_{k-2}(n,k;n,k) &=& \frac{(n+k)!}{n! k (k^2-1)} (2n+k+1) , \\[3pt]
K_{k-1}(n,k;n,k) &=& \frac{(n+k)!}{n! k} , \\[3pt]
K_{k+1}(n,k;n,k) &=& \frac{(n+k)!}{n!} (2n+k+1) , \\[3pt]
K_{k+2}(n,k;n,k) &=& \frac{(n+k)!}{n!} \Big \{ 6n(n+k+1) + (k+1)(k+2) \Big \} , \\[3pt]
K_{k+1}(n,k;n,k+1) &=& \frac{(n+k+1)!}{n!} , \\[3pt]
K_{k+1}(n,k;n-1,k+1) &=& - \frac{(n+k)!}{(n-1)!} , \\[3pt]
K_{k+2}(n,k;n-1,k+1) &=& - \frac{(n+k)!}{(n-1)!} (3n+2k+1) , \\[3pt]
K_0(n-1,1;n-a,a) &=& \binom{n}{a} \big \{ 1 - \delta_{a=0} \big \} = \frac{n!}{a! (n-a)!} \big \{ 1 - \delta_{a=0} \big \} \quad (a \in \mathbf{Z}, 0 \le a \le n) , \\[3pt]
K_0(n-2,3;n-2,3) &=& \frac{n (n^2-1) (3n^2-2)}{60} , \\[3pt]
K_0(n-2,3;n-3,4) &=& \frac{n (n^2-1) (n-2) (5n^2+n-3)}{360} , \\[3pt]
K_0(n-3,4;n-3,4) &=& \frac{n (n^2-1) (n-2) (2n-1) (5n^2-5n-9)}{2520} , \\[3pt]
K_1(n-1,1;n-a,a) &=& n \, \delta_{a=1} \quad (a \in \mathbf{Z}, 1 \le a \le n) , \\[3pt]
K_2(n-1,1;n-1,1) &=& 2n^2 , \\[3pt]
K_2(n-1,1;n-2,2) &=& -n(n-1) , \\[3pt]
K_3(n-1,1;n-3,5) &=& n (n-1) (n-2) , \\[3pt]
K_4(n-1,1;n-3,5) &=& 2 n (n-1) (n-2) (n+3) , \\[3pt]
K_4(n-2,2;n-3,5) &=& -n (n-1) (n-2) (n+5) .
\eea \ese
Some values for the subtracted $K$ integrals are
\bse \bea
{}^1 \! K_{-1}(n-1,1;n-1,1) &=& -\frac{n^{\underline{2}}}{2} = -\frac{n(n-1)}{2} , \\[3pt]
{}^1 \! K_{-1}(n-2,2;n-1,1) &=& -\frac{n^{\underline{3}}}{6} = -\frac{n(n-1)(n-2)}{6} , \\[3pt]
{}^2 \! K_{-1}(n-1,1;n-1,1) &=& 0 , \\[3pt]
{}^2 \! K_{-2}(n-1,1;n-1,1) &=& \frac{n^{\underline{3}}}{12} = \frac{n(n-1)(n-2)}{12} , \\[3pt]
{}^2 \! K_{-1}(n-1,1;n-2,2) &=& \frac{n^{\underline{3}}}{12} = \frac{n(n-1)(n-2)}{12} , \\[3pt]
{}^2 \! K_{1}(n-1,1;n-3,5) &=& \frac{n^{\underline{3}}}{3} (n-3/2) = \frac{1}{3} n (n-1) (n-2) (n-3/2) , \\[3pt]
{}^2 \! K_2(n-1,1;n-3,5) &=& 2 n^{\underline{3}} = 2n(n-1)(n-2) , \\[3pt]
{}^2 \! K_2(n-1,1;n-4,6) &=& \frac{7 n^{\underline{4}}}{6} = \frac{7}{6} n(n-1)(n-2)(n-3) .
\eea \ese

A general expression for the evaluation of ${}^p \! K_s(n,k;n',k')$ can be obtained by considering three possible regions in the sum over $r$.  Our general assumptions are that $n, k, n', k', p$ are all non-negative integers and that the convergence condition $s+p > -1$ holds.  While the general formula (\ref{Laguerre_Ktilde_int}) has no restriction that $s$ be an integer, we now assume that $s$ is an integer as well. We note that the factors
\be
\frac{(-1)^r \Gamma(s+r+1)}{r! (n-r)! (k+r)!}
\ee
in the sum over $r$ are all well defined and finite for $r$ in the summation range $p \le r \le n$.  For $p \le r \le k'-s-1$, both $\Gamma(s+r+1-k')$ in the numerator and $\Gamma(s+r+1-k'-n')$ in the denominator are divergent.  The ratio is evaluated as a limit in which $s \rightarrow \epsilon+s$ approaches its integral value $s$ when $\epsilon \rightarrow 0$.  One has
\be
\frac{\Gamma(s+r+1-k')}{\Gamma(s+r+1-k'-n')} \rightarrow \lim_{\epsilon \rightarrow 0} \frac{\Gamma(\epsilon-(k'-s-1-r))}{\Gamma(\epsilon-(n'+k'-s-1-r)}
= \frac{(-1)^{n'+k'-s-1-r} (n'+k'-s-1-r)!}{(-1)^{k'-s-1-r} (k'-s-1-r)!}
\ee
For values of $r$ in the range $k'-s \le r \le n'+k'-s-1$, the divergent gamma function in the denominator is not canceled by a numerator divergence and the contribution vanishes.  For $r$ in the range $n'+k'-s \le r \le n$, all gamma functions are finite and can be written as factorials.  The result for the subtracted $K$ integral is
\bea \label{general_K_integral}
{}^p \! K_s(n,k;n',k') &=& \frac{(n+k)! (-1)^{n'}}{n'!} \Bigg \{ \sum_{r=p}^{k'-s-1} \frac{(-1)^r (r+s)! (-1)^{n'} (n'+k'-s-1-r)!}{r! (r+k)! (n-r)! (k'-s-1-r)!} \crr
&\hbox{}& \hskip 2.4cm + \sum_{r=n'+k'-s}^{n} \frac{(-1)^r (r+s)! (r+s-k')!}{r! (r+k)! (n-r)! (r+s-n'-k')!} \Bigg \} .
\eea
This formula works as well when $p=0$, giving the standard $K$ integrals of (\ref{gen_Laguerre_int}).  We note that one or both of the summation regions might be void, giving a contribution of zero for that sum.  Also, the factorial functions in the denominators enforce the overall limits of summation, keeping $r$ in the range $0 \le r \le n$.

An associated integral with a log function can be obtained by way of the same method of differentiation by $s$ that gave $J_s(n,k)$ from $I_s(n,k)$:
\bea
L_s(n,k;n',k') &=& \int_0^\infty dx \, e^{-x} x^s \ln x \, L_n^k(x) L_{n'}^{k'}(x) \crr
&=& \frac{(n+k)! (-1)^{n'}}{n'!} \sum_{r=0}^n \frac{(-1)^r \Gamma(s+r+1-k') \Gamma(s+r+1)}{r! (n-r)! (k+r)! \Gamma(s+r+1-k'-n')} \crr
&\hbox{}& \quad \times \Bigl \{ \psi(s+r+1-k') + \psi(s+r+1) - \psi(s+r+1-k'-n') \Bigr \} .
\eea
Some examples include:
\bse \bea
\label{integral_L0nm11nm11}
L_0(n-1,1;n-1,1) &=& -n(\text{H}_n + \gamma_E - 1) , \\[3pt]
\label{integral_L1nm11nm11}
L_1(n-1,1;n-1,1) &=& n(\text{H}_n - \gamma_E ) , \\[3pt]
\label{integrat_L2nm11nm11}
L_2(n-1,1;n-1,1) &=& 2n^2 (\text{H}_n-\gamma_E+1)-n , \\[3pt]
\label{integral_L2nm11nm22}
L_2(n-1,1;n-2,2) &=& -n(n-1) (\text{H}_n+1-\gamma_E) , \\[3pt]
L_{k-2}(n,k;n,k) &=& \frac{(n+k)!}{n! k (k^2-1)} \big [ (2n+k+1) \big ( \text{H}_{k+1}+\text{H}_{k-2}-\text{H}_{n+k}-\gamma_E \big ) - (2n+1) \big ] , \\[3pt]
L_{k-1}(n,k;n,k) &=& \frac{(n+k)!}{n! k} \bigl [ 2\text{H}_k-\text{H}_{n+k}-\frac{1}{k}-\gamma_E \bigr ] , \\[3pt]
\label{integral_Lknknk}
L_k(n,k;n,k) &=& \frac{(n+k)!}{n!} (\text{H}_{n+k}-\gamma_E) , \\[3pt]
\label{integral_Lkp1nknk}
L_{k+1}(n,k;n,k) &=& \frac{(n+k)!}{n!} \bigl [ (2n+k+1)(\text{H}_{n+k}-\gamma_E)+2n+1 \bigr ] , \\[3pt]
L_{k+1}(n,k;n-1,k+1) &=& - \frac{(n+k)!}{(n-1)!} (\text{H}_{n+k}-\gamma_E+1)  .
\eea \ese

The subtracted $L$ integral is defined as
\bea
{}^p \! L_s(n,k;n',k') &=& \int_0^\infty dx \, e^{-x} x^s \ln x \, ^p \! L_n^k(x) L_{n'}^{k'}(x) \crr
&=& \frac{(n+k)! (-1)^{n'}}{n'!} \sum_{r=p}^n \frac{(-1)^r \Gamma(s+r+1-k') \Gamma(s+r+1)}{r! (n-r)! (k+r)! \Gamma(s+r+1-k'-n')} \crr
&\hbox{}& \quad \times \Bigl \{ \psi(s+r+1-k') + \psi(s+r+1) - \psi(s+r+1-k'-n') \Bigr \} ,
\eea
where $\psi(n,z)$ is defined in (\ref{eqn_def_psi_n_z}).  As an example, one has
\be
{}^2 \! L_1(n-1,1;n-3,5) = -\frac{1}{72}(n-1)(n-2) \Bigl \{ 36+75n-73n^2 + 24 n (n-3/2) \big ( \text{H}_n+\gamma_E \big ) \Bigr \} .
\ee

A second differentiation by $s$ gives integrals involving $\ln^2 x$:
\be
\frac{d^2}{ds^2} x^s = \ln^2 x \; x^s .
\ee
We find that
\bea
M_s(n,k;n',k') &=& \int_0^\infty dx \, e^{-x} x^s \ln^2 x \, L_n^k(x) L_{n'}^{k'}(x) \crr
&=& \frac{(n+k)! (-1)^{n'}}{n'!} \sum_{r=0}^n \frac{(-1)^r \Gamma(s+r+1-k') \Gamma(s+r+1)}{r! (n-r)! (k+r)! \Gamma(s+r+1-k'-n')} \crr
&\hbox{}& \quad \times \Bigl \{ \psi(1,s+r+1-k') + \psi(1,s+r+1) - \psi(1,s+r+1-k'-n') \crr
&\hbox{}& \quad \; \; \; + \psi^2(s+r+1-k') + \psi^2(s+r+1) + \psi^2(s+r+1-k'-n') \crr
&\hbox{}& \quad \; \; \; + 2 \psi(s+r+1-k') \psi(s+r+1) - 2 \psi(s+r+1-k') \psi(s+r+1-k'-n') \crr
&\hbox{}& \quad \; \; \; - 2 \psi(s+r+1) \psi(s+r+1-k'-n') \Bigr \} .
\eea
As an example, we evaluate $M_1(n-1,1;n-1,1)$.  Firstly, by explicit integration, one has the values $\gamma_E^2 - 2 \gamma_E+ \zeta(2)$ and $2 \gamma_E^2 - 6 \gamma_E + 2 \zeta(2) + 4$ for $n=1$ and $n=2$ respectively.  In general one has
\bea
M_1(n-1,1;n-1,1) &=& \int_0^\infty dx \, e^{-x} x \ln^2 x \, L_{n-1}^1(x) L_{n-1}^1(x) \crr
&=& n (-1)^{n-1} \sum_{r=0}^{n-1} \frac{(-1)^r }{(n-1-r)! \Gamma(r+2-n)} 
\Bigl \{ \psi(1,r+1) + \psi(1,r+2)  \crr
&\hbox{}& \quad \; \; \; - \psi(1,r+2-n) + \psi^2(r+1) + \psi^2(r+2) + \psi^2(r+2-n) \crr
&\hbox{}& \quad \; \; \; + 2 \psi(r+1) \psi(r+2) - 2 \psi(r+1) \psi(r+2-n) - 2 \psi(r+2) \psi(r+2-n) \Bigr \} .
\eea
For general $n \ge 2$ one breaks the sum into two parts: $0 \le r \le n-2$ and $r=n-1$ and works them out separately.  The gamma function in the denominator is singular in the sum, so only the singular parts of the numerator contribute and a limiting process is required for their evaluation.  The result is
\bea
M_1(n-1,1;n-1,1) &=& n \Bigl \{ \zeta(2,n) + \zeta(2,n+1) - \zeta(2,1) + (\text{H}_{n-1}+\text{H}_n)^2 -2 \gamma_E (\text{H}_{n-1} + \text{H}_n) + \gamma_E^2 \Bigr \} \crr
&\hbox{}& \quad - 2 n \sum_{r=0}^{n-2} \frac{1}{n-1-r} \Bigl \{ \text{H}_r + \text{H}_{r+1} - \gamma_E - \text{H}_{n-2-r} \Bigr \} ,
\eea
where the sum is void for $n=1$ and $\text{H}_0 \equiv 0$.  The sums can be done using the formulas of Appendix \ref{special_functions}, resulting in
\be
M_1(n-1,1;n-1,1) = n \left \{ \text{H}_{n-1}^2 + \text{H}_{n-1}^{(2)} + \zeta(2) - 2 \gamma_E \text{H}_n + \gamma_E^2 \right \} .
\ee
More generally, one has
\be
M_k(n,k;n,k) = \frac{(n+k)!}{n!} \biggl \{ \text{H}_{n+k}^2 - \text{H}_{n+k}^{(2)} + 2 \text{H}_{n+k} \left ( \text{H}_n - \gamma_E \right ) - 2 \text{diH}_-(n,n+k-1) + \gamma_E^2 + \zeta(2) \biggr \} ,
\ee
where $\text{diH}_-(n,m)$ is the falling diharmonic number defined in (\ref{diharmonic_defs}).

\section{Discussion}
\label{discussion}

We have given an extensive list of expectation values that are useful for the calculation of corrections to the properties of Coulombic two-body bound systems.  We used dimensional regularization to regulate divergences and have found the values for a number of expectation values using this regulatory scheme.  In addition to exploring the use of dimensional regularization in the Coulomb problem, we gave convenient formulas for a number of integrals involving the standard associated Laguerre polynomials.  These integration formulas involved one or two associated Laguerre polynomials, arbitrary powers of the radial variable $r$, and possibly various powers of $\ln r$.  We also obtained formulas for analogous integrals involving subtracted associated Laguerre polynomials that were required for the evaluation of divergent expectation values.  Most of the tabulated expectation values, Laguerre polynomial integrals, and associated sums described in Sec.~\ref{Laguerre_integrals} were used in a calculation of corrections to the energy levels of positronium at order $m \alpha^6$ using dimensionally regularized nonrelativistic quantum electrodynamics (NRQED) \cite{Adkins18a,Adkins19}, and would also appear in calculations of energy levels for any Coulombic two-body bound state at this order when using dimensional regularization.

\vfill \break

\appendix

\section{Coulomb expectation values}
\label{expectation_values}

We define coordinate space expectation values in the usual way:
\be
 \left \langle M \right \rangle = \int d^D x \, \psi^\dagger(\vec x\,) M(\vec x\,) \psi(\vec x\,) .
\ee

Some finite expectation values in $D=3$ dimensions are given here:
\bse \bearray
\left \langle \delta^3(\vec x\,) \right \rangle &=& \vert \psi(\vec x = 0) \vert^2 = \phi_n^2 = \frac{\gamma_n^3}{\pi} = \frac{(m_r Z \alpha)^3}{\pi n^3}, \label{onia} \\[3pt]
\left \langle 1 \right \rangle &=& 1 , \\[3pt]
\bigl \langle r \bigr \rangle &=& \frac{1}{2} \Bigl \{ 3n^2-\ell(\ell+1) \Bigr \} (m_r Z \alpha)^{-1} , \\[3pt]
\bigl \langle r^2 \bigr \rangle &=& \frac{n^2}{2} \Bigl \{ 5 n^2 + 1 - 3 \ell(\ell+1) \Bigr \} (m_r Z \alpha)^{-2} , \\[3pt]
\bigl \langle r^3 \bigr \rangle &=& \frac{n^2}{8} \Bigl \{ 35n^4 + 25 n^2 - 30 n^2 \ell(\ell+1) - 6 \ell(\ell+1) + 3 \bigl [ \ell(\ell+1) \bigr ]^2 \Bigr \} (m_r Z \alpha)^{-3} , \\[3pt]
\bigl \langle r^4 \bigr \rangle &=& \frac{n^4}{8} \Bigl \{ 63n^4 + 105 n^2 +12 - 70 n^2 \ell(\ell+1) - 50 \ell(\ell+1) + 15 \bigl [ \ell(\ell+1) \bigr]^2 \Bigr \} (m_r Z \alpha)^{-4} , \\
\left \langle \frac{1}{r} \right \rangle &=& \frac{m_r Z \alpha}{n^2} , \label{rminus1} \\[3pt]
\left \langle \frac{1}{r^2} \right \rangle &=& \frac{(m_r Z \alpha)^2}{(\ell+1/2) n^3} , \label{rminus2} \\[3pt]
\left \langle \frac{1}{r^3} \right \rangle_{\ell>0} &=& \frac{(m_r Z \alpha)^3}{\ell (\ell+1) (\ell+1/2) n^3} , \label{rminus3} \\[3pt]
\left \langle \frac{1}{r^4} \right \rangle_{\ell>0} &=& \frac{[3n^2-\ell (\ell+1)] (m_r Z \alpha)^4}{2 \ell (\ell+1) (\ell+1/2) (\ell-1/2) (\ell+3/2) n^5} , \label{rminus4} \\[3pt]
\left \langle \frac{1}{r^5} \right \rangle_{\ell>1} &=& \frac{\left ( 5n^2+1-3 \ell(\ell+1) \right ) (m_r Z \alpha)^5}{2 (\ell-1) \ell (\ell+1) (\ell+2) (\ell-1/2) (\ell+1/2) (\ell+3/2) n^5}  , \label{rminus5} \\[3pt]
\llangle \frac{1}{r^2} \delta^3(\vec x\,) \rrangle_{\ell>0} &=& \frac{(n^2-1)}{9 \pi n^5} (m_r Z \alpha)^5 \delta_{\ell=1} .
\eea \ese
A set of finite expectation values containing the momentum operator $\vec p = - i \vec \nabla$ follows:
\bse \bea
\left \langle p^2 \right \rangle &=& \gamma_n^2 = \frac{(m_r Z \alpha)^2}{n^2} , \label{p2} \\[3pt]
\left \langle p^4 \right \rangle &=& \left \{ \frac{4}{(\ell+1/2)n^3} - \frac{3}{n^4} \right \} (m_r Z \alpha)^4 , \label{p4} \\[3pt]
\label{expec_p6}
\left \langle p^6 \right \rangle &=& \biggl \{ \frac{5}{n^6} - \frac{8}{(\ell+1/2) n^5} + \frac{8n^2+1-4\ell(\ell+1)}{(\ell-1/2)(\ell+1/2)(\ell+3/2)n^5} + \frac{32}{n^3} \delta_{\ell=0} \biggr \} (m_r Z \alpha)^6 , \crr \\[3pt]
\left \langle p_i \frac{1}{r} p_i \right \rangle 
&=&  \left \{ \frac{2}{(\ell+1/2)n^3} - \frac{1}{n^4} - \frac{2}{n^3} \delta_{\ell=0} \right \} (m_r Z \alpha)^3 , \\[3pt]
\left \langle p_i \frac{1}{r^2} p_i \right \rangle 
&=& \left \{ \frac{8n^2+1-4\ell(\ell+1)}{4(\ell-1/2)(\ell+1/2)(\ell+3/2) n^5} + \frac{8}{n^3} \delta_{\ell=0} \right \} (m_r Z \alpha)^4 \label{p_rminus2_p} , \\[3pt]
\blangle p_i \delta^3(\vec x\,) p_i \brangle &=& \left \{ \frac{\delta_{\ell=0}}{n^3} + \frac{(n^2-1) \delta_{\ell=1}}{3 n^5} \right \} \frac{(m_r Z \alpha)^5}{\pi} , \\[3pt]
\blangle p_i \hat x_i \hat x_j p_j \brangle &=& \left \{ \frac{1}{n^2} - \frac{\ell(\ell+1)}{(\ell+1/2)n^3} \right \} ( m_r Z \alpha)^2 , \\[3pt]
\llangle p_i \hat x_i \frac{1}{r} \hat x_j p_j \rrangle &=& \left \{ - \frac{1}{n^4} + \frac{1}{(\ell+1/2)n^3} \right \} ( m_r Z \alpha)^3 , \\[3pt]
\llangle p_i \hat x_i \frac{1}{r^2} \hat x_j p_j \rrangle &=& \left \{ \frac{2n^2+1-2\ell(\ell+1)}{4(\ell-1/2)(\ell+1/2)(\ell+3/2) n^5} + \frac{4}{n^3} \delta_{\ell=0} \right \} (m_r Z \alpha)^4 , \\[3pt]
\llangle p_i \frac{1}{r} p_i - 3 p_i \hat x_i \frac{1}{r} \hat x_j p_j \rrangle &=& \left \{ \frac{2}{n^4} - \frac{1}{(\ell+1/2)n^3} - \frac{2}{n^3} \delta_{\ell=0} \right \} (m_r Z \alpha)^3 , \\[3pt]
\llangle p_i \frac{1}{r^2} p_i - 3 p_i \hat x_i \frac{1}{r^2} \hat x_j p_j \rrangle &=& \left \{ \frac{n^2-1+\ell(\ell+1) }{2(\ell-1/2)(\ell+1/2)(\ell+3/2) n^5} - \frac{4}{n^3} \delta_{\ell=0} \right \} (m_r Z \alpha)^4 , \\[3pt]
\llangle p_i \frac{1}{r^3} p_i - 3 p_i \hat x_i \frac{1}{r^3} \hat x_j p_j \rrangle_{\ell>0}  &=& \left \{ \frac{3n^2-\ell(\ell+1)}{2\ell(\ell+1)(\ell-1/2)(\ell+1/2)(\ell+3/2)n^5} + \frac{2(n^2-1)}{9n^5} \delta_{\ell=1} \right \} (m_r Z \alpha)^5 , \\[3pt]
\llangle p^2 r \rrangle &=& \llangle r p^2 \rrangle = \left \{ \frac{\ell(\ell+1)}{2n^2} + \frac{1}{2} \right \} (m_r Z \alpha) , \\[3pt]
\left \langle p^2 \frac{1}{r} \right \rangle &=& \left \langle \frac{1}{r} p^2 \right \rangle = \left \{ \frac{2}{(\ell+1/2) n^3} - \frac{1}{n^4} \right \} (m_r Z \alpha)^3 , \\[3pt]
\left \langle p^4 \frac{1}{r} \right \rangle_{\ell>0} &=& \left \langle \frac{1}{r} p^4 \right \rangle_{\ell>0} = \left \{ \frac{4n^2+2-4\ell(\ell+1)}{(\ell-1/2)(\ell+1/2)(\ell+3/2)n^5} + \frac{1}{n^6} \right \} (m_r Z \alpha)^5 ,\\[3pt]
\left \langle p^2 \frac{1}{r^2} \right \rangle_{\ell>0} &=& \left \langle \frac{1}{r^2} p^2 \right \rangle_{\ell>0} = \left \{ \frac{2}{\ell(\ell+1)(\ell+1/2)n^3} - \frac{1}{(\ell+1/2) n^5} \right \} (m_r Z \alpha)^4 , \\[3pt]
\left \langle p^2 \frac{1}{r^3} \right \rangle_{\ell>0} &=& \left \langle \frac{1}{r^3} p^2 \right \rangle_{\ell>0} = \left \{ \frac{3n^2+3/4-2\ell(\ell+1)}{\ell(\ell+1)(\ell+1/2)(\ell-1/2)(\ell+3/2) n^5} \right \} (m_r Z \alpha)^5 , \\[3pt]
\llangle p^2 r p^2 \rrangle &=& \left \{ - \frac{\ell(\ell+1)}{2n^4} + \frac{3}{2n^2} \right \} (m_r Z \alpha)^3 , \\[3pt]
\llangle p^2 \frac{1}{r} p^2 \rrangle_{\ell>0} &=& \left \{ \frac{1}{n^6} + \frac{4n^2-4\ell(\ell+1)}{\ell(\ell+1)(\ell+1/2)n^5} \right \} (m_r Z \alpha)^5 .
\eearray \ese
Expectation values involving the radial derivative $\partial_r = \hat x_i \partial_i = i \hat x \cdot \vec p$ include
\bse \bearray
\bigl \langle r \partial_r \bigr \rangle &=& - \frac{3}{2} , \\[3pt]
\bigl \langle \partial_r \bigr \rangle &=& - \frac{m_r Z \alpha}{n^2} , \\[3pt]
\left \langle \frac{1}{r} \partial_r \right \rangle &=& - \frac{(m_r Z \alpha)^2}{2(\ell+1/2) n^3} , \\[3pt]
\left \langle \frac{1}{r^2} \partial_r \right \rangle &=& - \frac{2 (m_r Z \alpha)^3}{n^3} \delta_{\ell=0} , \\[3pt]
\left \langle \frac{1}{r^3} \partial_r \right \rangle_{\ell>0} &=& \frac{\left ( 3n^2-\ell(\ell+1) \right ) (m_r Z \alpha)^4}{4 \ell (\ell+1) (\ell-1/2) (\ell+1/2) (\ell+3/2) n^5} , \\[3pt]
\left \langle \frac{1}{r^3} \Bigl ( \partial_r + m_r Z \alpha \Bigr ) \right \rangle &=& \left \{ \frac{4n^2-1}{4(\ell-1/2)(\ell+1/2)(\ell+3/2) n^5} + \frac{2}{n^3} \delta_{\ell=0} \right \} (m_r Z \alpha)^4 , \\[3pt]
\left \langle \frac{1}{r^4} \partial_r \right \rangle_{\ell>1} &=& \frac{\left ( 5n^2+1 -3 \ell(\ell+1) \right ) (m_r Z \alpha)^5}{2 (\ell-1) \, \ell \, (\ell+1) (\ell+2) (\ell-1/2) (\ell+1/2) (\ell+3/2) n^5} , \\[3pt]
\llangle \frac{1}{r^4} \left ( \partial_r - \frac{1}{r} \right ) \rrangle_{\ell > 0} &=& - \frac{2(n^2-1) (m_r Z \alpha)^5}{9 n^5} \delta_{\ell=1} .
\eearray \ese
Some expectation values involving $\partial_r^2$ are
\bse \bearray
\bigl \langle r \partial_r^2 \bigr \rangle &=& \left \{ \frac{4+\ell(\ell+1)}{2n^2} - \frac{1}{2}  \right \} (m_r Z \alpha), \\[3pt]
\bigl \langle \partial_r^2 \bigr \rangle &=& \left \{ \frac{1+\ell(\ell+1)}{(\ell+1/2)n^3} - \frac{1}{n^2}  \right \} (m_r Z \alpha)^2, \\[3pt]
\left \langle \frac{1}{r} \partial_r^2 \right \rangle &=& \left \{ \frac{1}{n^4}-\frac{1}{(\ell+1/2)n^3}+\frac{2}{n^3} \delta_{\ell=0} \right \} (m_r Z \alpha)^3, \\[3pt]
\label{expec_rm2dr2} 
\left \langle \frac{1}{r^2} \partial_r^2 \right \rangle 
&=& \frac{ \left ( -2n^2-1+2\ell(\ell+1) \right )(m_r Z \alpha)^4}{4(\ell-1/2)(\ell+1/2)(\ell+3/2) n^5} , \\[3pt]
\left \langle \frac{1}{r^3} \partial_r^2 \right \rangle_{\ell>0} 
&=& \left \{ \frac{-n^2-1/2+\ell(\ell+1)}{2\ell(\ell+1)(\ell-1/2)(\ell+1/2)(\ell+3/2) n^5} - \frac{2(n^2-1)}{9 n^5} \delta_{\ell=1} \right \} (m_r Z \alpha)^5 ,
\eearray \ese
and also
\bse \bea
\blangle \partial_r^\dagger \partial_r^2 \brangle &=& \left \{ \frac{1}{n^4} - \frac{1}{(\ell+1/2)n^3} \right \} (m_r Z \alpha)^3 , \\[3pt]
\llangle \partial_r^\dagger \frac{1}{r} \partial_r^2 \rrangle &=& \left \{ \frac{-n^2-1/2+\ell(\ell+1)}{4(\ell-1/2)(\ell+1/2)(\ell+3/2)n^5} - \frac{2}{n^3} \delta_{\ell=0} \right \} (m_r Z \alpha)^4 , \\[3pt]
\llangle \partial_r^\dagger \frac{1}{r^2} \partial_r^2 \rrangle &=& \left \{ -\frac{2}{n^3} \delta_{\ell=0} - \frac{2(n^2-1)}{9n^5} \delta_{\ell=1} \right \} (m_r Z \alpha)^5 , \\[3pt]
\blangle (\partial_r^\dagger)^2 \partial_r^2 \brangle &=& \left \{ \frac{-4n^2-2-2\ell(\ell+1)+6n^2\ell(\ell+1)+6(\ell(\ell+1))^2}{4(\ell-1/2)(\ell+1/2)(\ell+3/2)n^5} - \frac{3}{n^4} \right \} (m_r Z \alpha)^4 , \\[3pt]
\llangle p^2 r \partial_r^2 \rrangle &=& \left \{ - \frac{4+\ell(\ell+1)}{2n^4} + \frac{2+2\ell(\ell+1)}{(\ell+1/2)n^3} - \frac{3}{2n^2} \right \} (m_r Z \alpha)^3 , \\[3pt]
\llangle p^2 \partial_r^2 \rrangle &=& \left \{ - \frac{2n^2+1+\ell(\ell+1)}{(\ell+1/2) n^5} + \frac{3}{n^4} + \frac{4}{n^3} \delta_{\ell=0} \right \} (m_r Z \alpha)^4 .
\eea \ese
Some expectation values involving $\partial_r^3$ are
\bse \bea
\blangle \partial_r^3 \brangle &=&  \left \{ -\frac{3}{n^4} + \frac{3}{(\ell+1/2)n^3} - \frac{4}{n^3} \delta_{\ell=0} \right \} (m_r Z \alpha)^3 , \\[3pt]
\llangle \frac{1}{r} \partial_r^3 \rrangle &=& \left \{ \frac{3n^2+3/2-3\ell(\ell+1)}{4(\ell-1/2)(\ell+1/2)(\ell+3/2)n^5} + \frac{2}{n^3} \delta_{\ell=0} \right \} (m_r Z \alpha)^4 , \\[3pt]
\llangle \frac{1}{r^2} \partial_r^3 \rrangle &=& \left \{ - \frac{2(n^2+2)}{3n^5} \delta_{\ell=0} + \frac{2(n^2-1)}{9n^5} \delta_{\ell=1} \right \} (m_r Z \alpha)^5 , \\[3pt]
\blangle \partial_r^\dagger \partial_r^3 \brangle &=& \left \{ \frac{6n^2+3-6n^2\ell(\ell+1)-6(\ell(\ell+1))^2}{4(\ell-1/2)(\ell+1/2)(\ell+3/2)n^5} + \frac{3}{n^4} + \frac{4}{n^3} \delta_{\ell=0} \right \} (m_r Z \alpha)^4 .
\eea \ese
In the following we also use the adjoint radial derivative $\partial_r^\dagger = - \partial_i \hat x_i = -i \vec p \cdot \hat x$.  The general relation holds:
\be
\blangle (\partial_r^\dagger)^a f(\vec x\,) (\partial_r)^b \brangle = \blangle (\partial_r^\dagger)^b f^*(\vec x\,) (\partial_r)^a \brangle^* 
\ee
from which follows
\be
\blangle (\partial_r^\dagger)^a r^s \brangle = \blangle r^s (\partial_r)^a \brangle.
\ee
Some expectation values involving both $\partial_r^\dagger$ and $\partial_r$ are
\bse \bea
\bigl \langle p_i \hat x_i \hat x_j p_j \bigr \rangle = \bigl \langle \partial_r^\dagger \partial_r \bigr \rangle &=& \left \{ \frac{1}{n^2} - \frac{\ell(\ell+1)}{(\ell+1/2)n^3} \right \} ( m_r Z \alpha)^2 , \\[3pt]
\left \langle p_i \hat x_i \frac{1}{r} \hat x_j p_j \right \rangle = \left \langle \partial_r^\dagger \frac{1}{r} \partial_r \right \rangle &=& \left \{ - \frac{1}{n^4} + \frac{1}{(\ell+1/2)n^3} \right \} ( m_r Z \alpha)^3 , \\[3pt]
\left \langle p_i \hat x_i \frac{1}{r^2} \hat x_j p_j \right \rangle = \llangle \partial_r^\dagger \frac{1}{r^2} \partial_r \rrangle &=& \left \{ \frac{2n^2+1-2 \ell (\ell+1)}{4(\ell-1/2)(\ell+1/2)(\ell+3/2) n^5} + \frac{4}{n^3} \delta_{\ell=0} \right \} (m_r Z \alpha)^4 , \\[3pt]
\left \langle p_i \hat x_i \frac{1}{r^3} \hat x_j p_j \right \rangle_{\ell>1} = \llangle \partial_r^\dagger \frac{1}{r^3} \partial_r \rrangle_{\ell>1} &=& \frac{6n^2+\ell(\ell+1)\big (2n^2-1-2\ell(\ell+1) \big )}{4(\ell-1) \, \ell \, (\ell+1)(\ell+2)(\ell-1/2)(\ell+1/2)(\ell+3/2)n^5} (m_r Z \alpha)^5 , \\[3pt]
\llangle \partial_r^\dagger \frac{1}{r^3} \left ( \partial_r-\frac{1}{r} \right ) \rrangle_{\ell>0} &=& \bigg \{ \frac{n^2+1/2-\ell(\ell+1)}{2\ell(\ell+1)(\ell-1/2)(\ell+1/2)(\ell+3/2)n^5} - \frac{2(n^2-1)}{9n^5} \delta_{\ell=1} \bigg \} (m_r Z \alpha)^5 , \cr
&& \\[3pt]
\llangle \left ( \partial_r^\dagger-\frac{1}{r} \right ) \frac{1}{r^3} \left ( \partial_r-\frac{1}{r} \right ) \rrangle_{\ell>0} &=& \bigg \{ \frac{ n^2+1/2-\ell (\ell+1) }{2\ell(\ell+1)(\ell-1/2)(\ell+1/2)(\ell+3/2) n^5} \bigg \} (m_r Z \alpha)^5 .
\eea \ese
We also find
\bse \bea
\llangle \partial_r^\dagger p^2 \partial_r \rrangle &=& \left \{ \frac{2n^2-2+2\ell (\ell+1)}{4(\ell-1/2)(\ell+1/2)(\ell+3/2)n^5}+\frac{\ell (\ell+1)}{(\ell+1/2)n^5}+\frac{2}{(\ell+1/2)n^3}-\frac{3}{n^4} \right \} (m_r Z \alpha)^4 , \\[3pt]
\llangle p_n \frac{1}{r} \partial_r p_n \rrangle &=& \llangle p_n \frac{i}{r} \hat x_i p_i p_n \rrangle = \left \{ \frac{-4n^2-1/2+2\ell (\ell+1)}{4(\ell-1/2)(\ell+1/2)(\ell+3/2)n^5} - \frac{4}{n^3} \delta_{\ell=0} \right \} (m_r Z \alpha)^4 , \\[3pt]
\blangle p_n p_i \hat x_i \hat x_j p_j p_n \brangle &=& \blangle \partial_n^\dagger \partial_r^\dagger \partial_r \partial_n \brangle = \left \{ \frac{2n^2-2+2\ell(\ell+1)}{4(\ell-1/2)(\ell+1/2)(\ell+3/2)n^5}+\frac{\ell(\ell+1)}{(\ell+1/2)n^5}+\frac{2}{(\ell+1/2)n^3}-\frac{3}{n^4} \right \} (m_r Z \alpha)^4 . \crr
\eea \ese

Some expectation values involving logs are
\bse \bea
\bigl \langle \ln(\kappa r) \bigr \rangle &=& \ln \left ( \frac{\kappa n}{2 m_r Z \alpha} \right ) + \text{H}_{n+\ell}-\gamma_E + 1 - \frac{2\ell+1}{2n} , \\
\llangle \frac{\ln(\kappa r)}{r} \rrangle &=& \frac{m_r Z \alpha}{n^2} \left \{ \ln \left ( \frac{\kappa n}{2 m_r Z \alpha} \right ) + \text{H}_{n+\ell}-\gamma_E \right \} , \\
\left \langle \frac{\ln (\kappa r)}{r^2} \right \rangle &=& \frac{(m_r Z \alpha)^2}{ (\ell+1/2) n^3} \Bigl \{ \ln \left ( \frac{\kappa n}{2 m_r Z \alpha} \right ) + \text{H}_{2\ell+1} + \text{H}_{2\ell} - \text{H}_{n+\ell} - \gamma_E \Bigr \} , \\
\left \langle \frac{\ln (\kappa r)}{r^3} \right \rangle_{\ell>0} &=& \frac{(m_r Z \alpha)^3}{\ell(\ell+1) (\ell+1/2) n^3} \Bigl \{ \ln \left ( \frac{\kappa n}{2 m_r Z \alpha} \right ) + \text{H}_{2\ell+2} + \text{H}_{2\ell-1} - \text{H}_{n+\ell} - \gamma_E - \frac{n-\ell-1/2}{n} \Bigr \} , \\
\left \langle \frac{\ln^2 (\kappa r)}{r} \right \rangle &=& \frac{(m_r Z \alpha)}{n^2} \Bigl \{ \ln^2 \left ( \frac{\kappa n}{2 m_r Z \alpha} \right ) + 2 \ln \left ( \frac{\kappa n}{2 m_r Z \alpha} \right ) \big ( \text{H}_{n+\ell} - \gamma_E \big ) + H^2_{n+\ell} - H^{(2)}_{n+\ell} \crr
&\hbox{}& \hspace{1.8cm}  + 2 \text{H}_{n+\ell} \big ( \text{H}_{n-\ell-1} - \gamma_E \big ) -  2 \text{diH}_-(n-\ell-1, n+\ell-1) + \gamma_E^2 + \zeta(2) \Bigr \} , \\
\llangle \ln(\kappa r) \partial_r \rrangle &=& - \frac{m_r Z \alpha}{n^2} \left \{ \ln \left ( \frac{\kappa n}{2 m_r Z \alpha} \right ) + \text{H}_{n+\ell}-\gamma_E + \half \right \} .
\eea \ese

\vfill \break 
Expectation values involving the Coulomb potential include the following, where we note that S state expectations involving $\bar V$ must be evaluated in $D$ dimensions:
\bse \bea
\langle V \rangle &=& -\frac{m_r (Z \alpha)^2}{n^2} , \\[3pt]
\langle V^2 \rangle &=& \frac{m_r^2 (Z \alpha)^4}{(\ell+1/2) n^3} , \\[3pt]
\label{expectation_V3_table}
\langle \bar V^3 \rangle &=& \bcases \pi \bar \phi_n^2 (Z \alpha)^3 \bar \mu^{2 \epsilon} 
\left \{ - \frac{1}{\epsilon} - 4 \ln \left ( \frac{\mu n}{2 m_r Z \alpha} \right ) + 4 \text{H}_n - \frac{2}{n} - 4 \right \} & \text{if } \ell = 0 \\[3pt]
\frac{-m_r^3 (Z \alpha)^6}{\ell (\ell+1) (\ell+1/2) n^3} & \text{if } \ell > 0 \ecases , \\[3pt]
\langle \bar V \bar V' \rangle &=& \bcases \pi \bar \phi_n^2 (Z \alpha)^2 \bar \mu^{2 \epsilon} 
\left \{ - \frac{2}{\epsilon} - 4 \ln \left ( \frac{\mu n}{2 m_r Z \alpha} \right ) + 4 \text{H}_n - \frac{2}{n} - 2 \right \} & \text{if } \ell = 0 \\[3pt]
\frac{-m_r^3 (Z \alpha)^5}{\ell (\ell+1) (\ell+1/2) n^3} & \text{if } \ell > 0 \ecases , \\[3pt]
\label{expec_Vp2} 
\langle (\bar V')^2 \rangle &=& \bcases \pi \bar \phi_n^2 m_r (Z \alpha)^3 \bar \mu^{2 \epsilon}
\left \{ - \frac{2}{\epsilon} - 8 \ln \left ( \frac{\mu n}{2 m_r Z \alpha} \right ) + 8 \text{H}_n + \frac{4}{3 n^2} - \frac{4}{n} - \frac{16}{3} \right \} & \text{if } \ell = 0
 \\[3pt]
\frac{3n^2-\ell(\ell+1)}{2 \ell (\ell+1) (\ell-1/2) (\ell+1/2) (\ell+3/2) n^5} m_r^4 (Z \alpha)^6 & \text{if } \ell > 0 \ecases , \\[3pt]
\blangle (\bar V')^2 \brangle &=& 2 m_r \blangle \bar V^3 \brangle 
+ \left \{ \frac{4n^2-1}{2 (\ell-1/2)(\ell+1/2)(\ell+3/2)n^5} + \frac{8}{n^3} \delta_{\ell=0} \right \} m_r^4 (Z \alpha)^6 , \\[3pt]
\blangle \bar V^2 \partial_r^2 \brangle &=& \left \{ \frac{ -2n^2-1+2\ell(\ell+1) }{4(\ell-1/2)(\ell+1/2)(\ell+3/2) n^5} - \frac{2}{n^3} \delta_{\ell=0} \right \} m_r^4 (Z \alpha)^6 , \\[3pt]
\left \langle p_i V p_i \right \rangle 
&=& \left \{ \frac{1}{n^4} - \frac{2}{(\ell+1/2)n^3} + \frac{2}{n^3} \delta_{\ell=0} \right \} m_r^3  (Z \alpha)^4 , \\[3pt]
\langle \bar V p^2 \bar V \rangle &=& \langle p_i \bar V^2 p_i \rangle = 
\left \{ \frac{8n^2+1-4\ell(\ell+1)}{4(\ell-1/2)(\ell+1/2)(\ell+3/2)n^5} + \frac{8}{n^3} \delta_{\ell=0} \right \} m_r^4 (Z \alpha)^6 , \label{expec_V_p2_V} \\[3pt]
\blangle p^2 V \brangle &=& \blangle V p^2 \brangle = \left \{ \frac{1}{n^4} - \frac{2}{(\ell+1/2)n^3} \right \} m_r^3 (Z \alpha)^4 , \\[3pt]
\label{expec_p4barV}
\blangle p^4 \bar V \brangle &=& \blangle \bar V p^4 \brangle = \left \{ \frac{-4n^2-2+4\ell(\ell+1)}{(\ell-1/2)(\ell+1/2)(\ell+3/2)n^5} - \frac{1}{n^6} - \frac{16}{n^3} \delta_{\ell=0} \right \} m_r^5 (Z \alpha)^6 ,\\[3pt]
\blangle \bar V^2 p^2 \brangle = \blangle p^2 \bar V^2 \brangle
&=& \bcases \pi \bar \phi_n^2 m_r (Z \alpha)^3 \bar \mu^{2 \epsilon} 
\left \{ \frac{2}{\epsilon} + 8 \ln \left ( \frac{\mu n}{2 m_r Z \alpha} \right ) - 8 \text{H}_n + \frac{4}{n} - \frac{2}{n^2} + 8 \right \} & \text{if } \ell = 0 \\[3pt]
\frac{2n^2-\ell(\ell+1)}{\ell (\ell+1) (\ell+1/2) n^5} m_r^4 (Z \alpha)^6 & \text{if } \ell > 0 \ecases , \\[3pt]
\blangle p_i \bar V p_i \bar V \brangle &=& \left \{ \frac{4n^2+2-4\ell(\ell+1)}{4(\ell-1/2)(\ell+1/2)(\ell+3/2)n^5} + \frac{4}{n^3} \delta_{\ell=0} \right \} m_r^4 (Z \alpha)^6 - m_r \blangle \bar V^3 \brangle , \\[3pt]
\blangle \partial_r^\dagger \bar V \partial_r \bar V \brangle &=& \left \{ \frac{2n^2+1-2\ell(\ell+1)}{4(\ell-1/2)(\ell+1/2)(\ell+3/2)n^5} + \frac{4}{n^3} \delta_{\ell=0} \right \} m_r^4 (Z \alpha)^6 - \half \blangle (\bar V')^2 \brangle , \\[3pt]
\label{expec_p2barVp2}
\blangle p^2 \bar V p^2 \brangle
&=& \bcases 4 \pi \bar \phi_n^2 m_r ^2 (Z \alpha)^3 \bar \mu^{2 \epsilon} 
\left \{ - \frac{1}{\epsilon} - 4 \ln \left ( \frac{\mu n}{2 m_r Z \alpha} \right ) + 4 \text{H}_n - \frac{2}{n} - \frac{1}{4n^3} + \frac{2}{n^2} - 4 \right \} & \text{if } \ell = 0 \crr
\left ( - \frac{1}{n^6} + \frac{4}{(\ell+1/2)n^5} - \frac{4}{\ell (\ell+1) (\ell+1/2) n^3} \right ) m_r^5 (Z \alpha)^6 & \text{if } \ell > 0 \ecases . \\
\eea \ese
Only the leading terms are shown for S state expectation values involving $\bar V$--terms of $O(\epsilon)$ are ignored.  All expectation values for $\ell>0$ were evaluated in the $D \rightarrow 3$ limit and the results given are exact in that limit, as are expectation values involving the three-dimensional Coulomb potential $V$.

\vfill \break

Second and higher order derivatives in coordinate space expectation values can lead to unexpected results.  Some examples of expectation values that do or might seem to lead to unexpected results are shown here:
\bse \bea
\llangle \frac{1}{r^{2-4\epsilon}} \partial_r^2 \rrangle &=& (m_r Z \alpha)^4 \left \{ \frac{ -2n^2-1+2\ell(\ell+1) }{4(\ell-1/2)(\ell+1/2)(\ell+3/2) n^5} - \frac{2}{n^3} \delta_{\ell=0} \right \}  \ne \llangle \frac{1}{r^2} \partial_r^2 \rrangle , \\[3pt]
\llangle \frac{1}{r^{2-4\epsilon}} p^2 \rrangle &=& (m_r Z \alpha)^4 \left \{ \frac{-1}{(\ell+1/2)n^5} \right \} - \frac{2m_r}{\beta^2} \blangle \bar V^3 \brangle , \\[3pt]
\llangle \frac{1}{r^{1-4\epsilon}} \partial_r^3 \rrangle &=& (m_r Z \alpha)^4 \left \{ \frac{ 3n^2+3/2-3\ell(\ell+1)}{4(\ell-1/2)(\ell+1/2)(\ell+3/2) n^5} + \frac{4}{n^3} \delta_{\ell=0} \right \} \ne \llangle \frac{1}{r} \partial_r^3 \rrangle , \\[3pt]
\llangle \frac{1}{r^{-4\epsilon}} \partial_r^2 \bar V \rrangle &=& m_r^3 (Z \alpha)^4 \left \{ - \frac{1}{n^4} + \frac{1}{(\ell+1/2)n^3} \right \} + \frac{2}{\beta^2} \blangle \bar V^3 \brangle , \\[3pt]
\llangle \frac{1}{r^{-4\epsilon}} \partial_r^2 p^2 \rrangle &=& (m_r Z \alpha)^4 \left \{ - \frac{1+\ell(\ell+1)}{(\ell+1/2)n^5} + \frac{3}{n^4} - \frac{2}{(\ell+1/2)n^3} \right \} - \frac{4 m_r}{\beta^2} \blangle \bar V^3 \brangle , \\[3pt]
\llangle \frac{1}{r^{-4\epsilon}} p^2 \bar V \rrangle &=& m_r^3 (Z \alpha)^4 \left \{ \frac{1}{n^4} - \frac{2}{(\ell+1/2)n^3} + \frac{4}{n^3} \delta_{\ell=0} \right \} \ne \blangle p^2 V \brangle , \\[3pt]
\llangle \frac{1}{r^{-4\epsilon}} p^4 \rrangle &=& (m_r Z \alpha)^4 \left \{ - \frac{3}{n^4} + \frac{4}{(\ell+1/2)n^3} - \frac{8}{n^3} \delta_{\ell=0} \right \} \ne \blangle p^4 \brangle, \\[3pt]
\llangle \frac{1}{r^{1-4\epsilon}} \partial_r \bar V \rrangle &=& \llangle \frac{1}{r^{1-4\epsilon}} \big \{ \bar V \partial_r + \bar V' \big \} \rrangle = - Z \alpha \llangle \frac{1}{r^2} \partial_r \rrangle - \frac{(1-2\epsilon)}{\beta^2} \blangle \bar V^3 \brangle = - \frac{1}{\beta^2} \blangle \bar V^3 \brangle , \\[3pt]
\llangle p_i \frac{1}{r^{-4\epsilon}} p_i \bar V \rrangle &=&  \blangle p^2 \bar V \brangle = m_r^3 (Z \alpha)^4 \left \{ \frac{1}{n^4} - \frac{2}{(\ell+1/2)n^3} \right \} , \\[3pt]
\llangle \frac{1}{r^{-4\epsilon}} p_i \bar V p_i \rrangle &=& \blangle p_i \bar V p_i \brangle = m_r^3 (Z \alpha)^4 \left \{ \frac{1}{n^4} - \frac{2}{(\ell+1/2)n^3} + \frac{2}{n^3} \delta_{\ell=0} \right \} , \\[3pt]
\llangle \frac{1}{r^{1-4\epsilon}} \partial_r p^2 \rrangle &=& (m_r Z \alpha)^4 \left \{ \frac{1}{2(\ell+1/2)n^5} \right \} + \frac{2m_r}{\beta^2} \blangle \bar V^3 \brangle .
\eea \ese
All of these formulas are correct only through $O(\epsilon^0)$--there are uncalculated $O(\epsilon)$ corrections.

Some exact $D$-dimensional relations among expectation values involving the potential $\bar V=-\beta/r^{1-2\epsilon}$ where $\beta=\Gamma(1/2-\epsilon) Z \alpha \bar \mu^{2\epsilon} \pi^{-1/2+\epsilon}$ include (\ref{eqn_recursion_for_m2p4eps}) and
\bse \bea
\blangle \bar V p^2 \bar V \brangle &=& \blangle p_i \bar V^2 p_i \brangle , \\[3pt]
\blangle (\bar V')^2 \brangle &=& -2 \blangle \bar V \bar V' \partial_r \brangle , \\[3pt]
\blangle (\bar V')^2 \brangle &=& 2 m_r \blangle \bar V^3 \brangle + \blangle \bar V p^2 \bar V \brangle - 2 m_r E \blangle \bar V^2 \brangle , \\[3pt]
\blangle (\bar V')^2 \brangle &=& \blangle \bar V p^2 \bar V \brangle - \blangle p^2 \bar V^2 \brangle  = \blangle \bar V p^2 \bar V \brangle - \blangle \bar V^2 p^2 \brangle, \\[3pt]
\label{identity_Vbarp2}
2 m_r \blangle (\bar V')^2 \brangle &=& \blangle p^2 \bar V p^2 \brangle - \blangle p^4 \bar V \brangle = \blangle p^2 \bar V p^2 \brangle - \blangle \bar V p^4 \brangle , \\[3pt]
\label{eqn_expec_Vbarp_dr}
\blangle \bar V' \partial_r \brangle &=& -2 \pi \bar \phi^2_n Z \alpha \bar \mu^{2\epsilon} \delta_{\ell=0} .
\eea \ese

\vfill \break

Some momentum-space brackets involving $\vec q = \vec p_2 - \vec p_1$ are
\be \label{bracket_qm1}
\left \langle \frac{1}{\left \vert \vec q \, \right \vert} \right \rangle_{\vec p_2,\vec p_1}  = \frac{1}{2 \pi^2} \left \langle \frac{1}{r^2} \right \rangle
= \frac{ (m_r Z \alpha)^2}{2 \pi^2 (\ell+1/2) n^3} ,
\ee
\be
\left \langle \frac{1}{\vec q \, ^2} \right \rangle_{\vec p_2,\vec p_1} = \frac{1}{4 \pi} \left \langle \frac{1}{r} \right \rangle = \frac{m_r Z \alpha}{4 \pi n^2} ,
\ee
\be
\llangle \frac{1}{\vec q \,^4} \rrangle_{\vec p_2,\vec p_1} = - \frac{1}{16\pi} \Bigl \{ 3n^2-\ell(\ell+1) \Bigr \} (m_r Z \alpha)^{-1} ,
\ee
\be
\langle \ln \left ( \vert \vec q \, \vert \right ) \rangle_{\vec p_2 , \vec p_1} 
= \bcases \left \{ \ln \left ( \frac{2 m_r Z \alpha}{n} \right ) + \text{H}_n + \frac{n-1}{2n} \right \} \frac{(m_r Z \alpha)^3}{\pi n^3} & \rm{if} \; \ell=0 \\[3pt]
\frac{-1}{4 \ell (\ell+1) (\ell+1/2)} \frac{(m_r Z \alpha)^3}{\pi n^3} & \rm{if} \; \ell > 0 \ecases ,
\ee
\be
\left \langle \frac{\ln (\vert \vec q \, \vert)}{\vec q \, ^2} \right \rangle_{\vec p_2,\vec p_1} = - \frac{1}{4 \pi} \left \langle \frac{\ln r + \gamma_E}{r} \right \rangle = \left \{ \ln \left ( \frac{2 m_r Z \alpha}{n} \right ) - \text{H}_{n+\ell} \right \} \frac{m_r Z \alpha}{4 \pi n^2} ,
\ee
\be \label{bracket_p2dotp1_1}
\Bigl \langle \vec p_2 \cdot \vec p_1 \Bigr \rangle_{\vec p_2, \vec p_1} =  \left \{\frac{(n^2-1) \delta_{\ell=1}}{3 n^5} \right \} \frac{(m_r Z \alpha)^5}{\pi} , 
\ee
\be \label{bracket_p2p1qm1_1}
\left \langle \frac{\vec p_2 \cdot \vec p_1}{\left \vert \vec q \, \right \vert} \right \rangle_{\vec p_2,\vec p_1}  
= \left \{ \frac{8n^2-4\ell(\ell+1)+1}{4(\ell+1/2)(\ell-1/2)(\ell+3/2) n^5} + \frac{8}{n^3} \delta_{\ell=0} \right \} \frac{ (m_r Z \alpha)^4}{2 \pi^2} ,
\ee
\bea \label{bracket_p2qqp1qm3_1}
\left \langle \frac{ (\vec p_2 \cdot \vec q \,) (\vec q \cdot \vec p_1\,) }{\vert \vec q \, \vert^3} \right \rangle_{\vec p_2,\vec p_1} 
&=& \left \{ \frac{ 4n^2-1}{4(\ell+1/2)(\ell-1/2)(\ell+3/2) n^5} \right \} \frac{ (m_r Z \alpha)^4}{2 \pi^2} ,
\eea
\be \label{bracket_p2p1qm2_1}
\left \langle \frac{\vec p_2 \cdot \vec p_1}{\vec q\,^2}\right \rangle_{\vec p_2,\vec p_1}  = \frac{1}{4 \pi} \left \langle p_i \frac{1}{r} p_i \right \rangle
= \left \{ \frac{2}{(\ell+1/2)n^3} - \frac{1}{n^4} - \frac{2}{n^3} \delta_{\ell=0} \right \} \frac{ (m_r Z \alpha)^3}{4 \pi} ,
\ee
\bea \label{bracket_p2qqp1qm4_1}
\left \langle \frac{ (\vec p_2 \cdot \vec q \,) (\vec q \cdot \vec p_1\,) }{\vec q\,^4} \right \rangle_{\vec p_2,\vec p_1}
= \frac{1}{8 \pi} \left \{ \left \langle p^k \frac{1}{r} p^k \right \rangle - \llangle p_r^\dagger \frac{1}{r} p_r \rrangle \right \}
&=& \left \{ \frac{1}{(\ell+1/2)n^3} - \frac{2}{n^3} \delta_{\ell=0} \right \} \frac{ (m_r Z \alpha)^3}{8 \pi} ,
\eea
\be \label{bracket_p2p1_by_q}
\left \langle \frac{\vsq{p_2} \vsq{p_1} - (\vec p_2 \cdot \vec p_1)^2}{\vec q\,^4} \right \rangle_{\vec p_2,\vec p_1} 
= \left \{ - \frac{1}{n} + \frac{3}{2(\ell+1/2)} - \delta_{\ell=0} \right \} \frac{(m_r Z \alpha)^3}{4 \pi n^3} ,
\ee
\be
\left \langle \frac{\left ( \vsq{p_2} - \vsq{p_1} \right )^2}{\vec q \,^2} \right \rangle_{\vec p_2, \vec p_1} = \frac{m_r}{\pi Z \alpha \bar \mu^{2\epsilon}} \blangle (\bar V')^2 \brangle  ,
\ee
\be
\llangle \frac{\vsq{p_2} \vsq{p_1}}{\vec q\,^2} \rrangle_{\vec p_2,\vec p_1} = - \frac{1}{4\pi Z \alpha \bar \mu^{2\epsilon}} \blangle p^2 \bar V p^2 \brangle ,
\ee
\bearray 
\llangle \frac{\vsq{p_2} \vsq{p_1} - (\vec p_2 \cdot \vec p_1\,)^2}{\vec q \,^2} \rrangle_{\vec p_2, \vec p_1}
&=& \bigg \{ - \frac{ \blangle (\bar V')^2 \brangle }{4 m_r^4 (Z \alpha)^6 \bar \mu^{2\epsilon}}  - \frac{1}{2 n^5} \delta_{\ell=0} + \frac{n^2-1}{6 n^5} \delta_{\ell=1} \bigg \} \frac{(m_r Z \alpha)^5}{\pi} .
\eearray
We can keep higher order terms if necessary, for example
\be
\bigl \langle \bar V(\vec q\,) \big \rangle_{\vec p_2,\vec p_1} = - \frac{m_r (Z \alpha)^2}{n^2} \left \{1 + \epsilon \left [ 4 \ln \left ( \frac{\mu n}{2 m_r Z \alpha} \right ) + 4 \text{H}_{n+\ell} + \frac{2}{n} - 2 \right ] + O(\epsilon^2) \right \} .
\ee

\section{The Fourier transform in $D$ dimensions}
\label{Fourier_transform}

The $D$-dimensional Fourier transform is given by \cite{Samko78}
\be \label{fourier_dim_D}
\int \dbar^D p \, e^{i \vec p \cdot \vec x} p^n Y^D_{\ell m}(\hat p\,) = \frac{i^\ell 2^n}{\pi^{D/2} r^{n+D}} \frac{\Gamma \left ( \frac{\ell+D+n}{2} \right )}{\Gamma \left ( \frac{\ell-n}{2} \right ) } Y^D_{\ell m}(\hat x) ,
\ee
where the $Y^D_{\ell m}$ are $D$-dimensional spherical harmonics of angular momentum $\ell$, while $m$ represents the indices necessary to distinguish one spherical harmonic with angular momentum $\ell$ from another.  Formula (\ref{fourier_dim_D}) is singular when the arguments of the gamma functions are non-positive integers.  The Fourier transform contains delta functions and their derivatives when $(\ell-n)/2$ is a non-positive integer and logs when $(\ell+D+n)/2$ is a non-positive integer.  The three-dimensional special case is standard and is discussed, for example, in \cite{Adkins16}.  Some useful transforms include
\bse \bea
\label{FT_0}
\int \dbar^D p \, e^{i \vec p \cdot \vec x} \frac{1}{p^\alpha} &=& \frac{\Gamma\left ( \frac{D-\alpha}{2} \right )}{2^\alpha \pi^{D/2} \Gamma \left ( \frac{\alpha}{2} \right ) r^{D-\alpha}} , \\[3pt]
\label{FT_1}
\int \dbar^D p \, e^{i \vec p \cdot \vec x} \frac{p_i}{p^\alpha} &=& i \frac{\Gamma\left ( \frac{D+2-\alpha}{2} \right )}{2^{\alpha-1} \pi^{D/2} \Gamma \left ( \frac{\alpha}{2} \right ) r^{D+1-\alpha}} \hat x_i , \\[3pt]
\label{FT_2}
\int \dbar^D p \, e^{i \vec p \cdot \vec x} \, \frac{p_i p_j}{p^\alpha} &=& \frac{\Gamma\left ( \frac{D+2-\alpha}{2} \right )}{2^{\alpha-1} \pi^{D/2} \Gamma \left ( \frac{\alpha}{2} \right ) r^{D+2-\alpha}} \Big \{ \delta_{i j} - (D+2-\alpha) \hat x_i \hat x_j \Big \} , \\[3pt]
\label{FT_3}
\int \dbar^D p \, e^{i \vec p \cdot \vec x} \, \frac{p_i p_j p_k}{p^\alpha} &=& i \frac{\Gamma\left ( \frac{D+4-\alpha}{2} \right )}{2^{\alpha-2} \pi^{D/2} \Gamma \left ( \frac{\alpha}{2} \right ) r^{D+3-\alpha}} \Big \{ \big ( \delta_{i j} \hat x_k + \delta_{j k} \hat x_i + \delta_{k i} \hat x_j \big ) - (D+4-\alpha) \hat x_i \hat x_j \hat x_k \Big \} , \\[3pt]
\label{FT_4}
\int \dbar^D p \, e^{i \vec p \cdot \vec x} \, \frac{p_i p_j p_k p_\ell}{p^\alpha} &=& \frac{\Gamma\left ( \frac{D+4-\alpha}{2} \right )}{2^{\alpha-2} \pi^{D/2} \Gamma \left ( \frac{\alpha}{2} \right ) r^{D+4-\alpha}} \Big \{ \big ( \delta_{i j} \delta_{k \ell} + \delta_{i k} \delta_{j \ell} + \delta_{i \ell} \delta_{j k} \big ) \crr
&\hbox{}& \hspace{0.4cm} - (D+4-\alpha) \big ( \delta_{i j} \hat x_k \hat x_\ell + \delta_{i k} \hat x_j \hat x_\ell + \delta_{i \ell} \hat x_j \hat x_k + \delta_{j k} \hat x_i \hat x_\ell + \delta_{j \ell} \hat x_i \hat x_k + \delta_{k \ell} \hat x_i \hat x_j \big ) \crr
&\hbox{}& \hspace{0.4cm} + (D+4-\alpha)(D+6-\alpha) \hat x_i \hat x_j \hat x_k \hat x_\ell \Big \} ,
\eea \ese
where $p = \vert \vec p\, \vert$.  The transforms (\ref{FT_0}) and (\ref{FT_1}) are immediate since the transformed functions $1/p^\alpha$ and $p_i/p^\alpha$ have $\ell=0$ and $\ell=1$, respectively, while $p_i p_j/p^\alpha$ is a combination of $\ell=2$ and $\ell=0$, $p_i p_j p_k/p^\alpha$ is a combination of $\ell=3$ and $\ell=1$, and $p_i p_j p_k p_\ell/p^\alpha$ is a combination of $\ell=4$, 2, and 0.

\section{Review of the bound-state Coulomb problem in three dimensions}
\label{Coulomb_in_3d}

In the appendix we assemble the main formulas relevant to the three-dimensional bound-state Coulomb problem.  We use the conventions for associated Laguerre polynomials found in \cite{Abramowitz72,Gradshteyn80,Galindo90,Arfken01,Olver10} and that are built into the computer algebra systems Mathematica \cite{Mathematica11.1} and MatLab \cite{MATLAB19}.  Other references, including many textbooks on quantum mechanics, use a variety of other conventions--a partial tabulation is given in Liboff \cite[Table~10.3]{Liboff98}.

The coordinate-space Schr\"odinger-Coulomb equation is
\be
\left \{ \frac{\vec p\,^2}{2m_r} + V(r) \right \} \psi(\vec x\,) = E \psi(\vec x\,) ,
\ee
where $\vec p = -i \vec \nabla$, $m_r=m_1 m_2/(m_1+m_2)$ is the reduced mass and $V(r)=-Z \alpha/r$.  For the bound state Coulomb wave functions we have
\be \label{Coulomb_wf} \psi_{n \ell m}(\vec r\,) = R_{n \ell}(r) Y_{\ell m}(\hat r) \ee
where
\be \label{radial_wf}
R_{n \ell}(r) = \left \{\frac{4 (n-\ell-1)!}{a^3 n^4 (n+\ell)!} \right \}^{1/2} \rho^\ell e^{-\rho/2} L_{n-\ell-1}^{2\ell+1}(\rho)
\ee
with $\rho = 2 r/(a n)$, and $a$, the Bohr radius for a system with positive charge $Z$ and reduced mass $m_r$, given by
\be 
a = \frac{\hbar}{m_r Z \alpha c} = \frac{1}{m_r Z \alpha}  
\ee
in our units.  For $Z=1$ (as for hydrogen) with the assumption of no recoil (nucleus of infinite mass, $m_r = m_e$), the usual Bohr radius is
\be 
a_{\infty, \mathrm{Bohr}} = \frac{\hbar}{m_e \alpha c} \approx 0.052\,918 \, nm .
\ee
We have separated out the short distance behavior (given by $\rho^\ell$) and the long distance behavior (controlled by the exponential $e^{-\rho/2}$).  As examples, the S state radial functions are
\be \label{Sradial_wf}
R_{n 0}(r) = \left \{\frac{4}{a^3 n^5} \right \}^{1/2} e^{-\rho/2} L_{n-1}^{1}(\rho) ,
\ee
and the P state radial functions are
\be \label{Pradial_wf}
R_{n 1}(r) = \left \{\frac{4}{a^3 n^5 (n^2-1)} \right \}^{1/2} \rho \, e^{-\rho/2} L_{n-2}^{3}(\rho) .
\ee
The radial Schr\"odinger equation is
\be
\left ( \partial_r^2 + \frac{2}{r} \partial_r - \frac{\ell(\ell+1)}{r^2} + \frac{2 m_r Z \alpha}{r} - \frac{(m_r Z \alpha)^2}{n^2} \right ) R_{n \ell}(r) = 0 ,
\ee
and in terms of the dimensionless radial variable $\rho$ it is
\be
\left ( \partial_\rho^2 + \frac{2}{\rho} \partial_\rho - \frac{\ell(\ell+1)}{\rho^2} + \frac{n}{\rho} - \frac{1}{4} \right ) R_{n \ell}(\rho) = 0 .
\ee

The solutions to the radial equation can be expressed in terms of associated Laguerre polynomials.  The Laguerre polynomials are the conventionally normalized solutions to the equation
\be \left [ x \frac{d^2}{dx^2} + (1-x) \frac{d}{dx} + n \right ] L_n(x) = 0 , \ee
and have the generating function
\be \frac{1}{1-z} \exp{\left [ -\frac{x z}{1-z}\right ]} = \sum_{n=0}^\infty z^n L_n(x), \quad \vert z \vert <1 , \ee
Rodrigues' formula
\be 
L_n(x) = \frac{e^x}{n!} \frac{d^n}{dx^n} \left ( x^n e^{-x} \right ) , 
\ee
recursion relations
\bse \bearray
(n+1) L_{n+1}(x) &=& (2n+1-x) L_n(x)-n L_{n-1}(x) , \\
x L'_n(x) &=& n \left [ L_n(x)-L_{n-1}(x) \right ] ,
\eearray \ese
and orthogonality relation
\be 
\int_0^\infty dx \, e^{-x} L_n(x) L_m(x) = \delta_{n m} . 
\ee
An explicit formula is
\be 
L_n(x) = n! \sum_{r=0}^n \frac{(-1)^r}{(n-r)! r!^2} x^r .
\ee
An integral formula for them can be derived easily by use of a Laplace transform to solve the defining differential equation:
\be
L_n(x) = \oint \frac{dt}{2\pi i} \frac{(1+t)^n}{t^{n+1}} e^{-x t} ,
\ee
where the integration contour circles the origin in a counterclockwise direction \cite{Merzbacher61}, and they can be expressed in terms of hypergeometric functions as
\be
L_n(x) = {_1}F_1(-n,1;x).
\ee

The associated Laguerre polynomials
\be
L_n^k(x) = (-1)^k \frac{d^k}{dx^k} L_{n+k}(x)
\ee
are the regular solutions to the equation
\be 
\left [ x \frac{d^2}{dx^2} + (k+1-x) \frac{d}{dx} + n \right ] L_n^k(x) = 0 . 
\ee
An explicit formula for them is
\be \label{eqn_assoc_Laguerre_series}
L_n^k(x) = \sum_{r=0}^n \frac{(-1)^r (n+k)!}{(n-r)! (k+r)! r!} x^r  = \sum_{r=0}^n \frac{(-x)^r}{r!} \binom{n+k}{n-r}  . 
\ee
We use the general definition of the binomial symbol
\be
\binom{a}{b} = \frac{\Gamma(a+1)}{\Gamma(a-b+1) \Gamma(b+1)} ,
\ee
which works even when $a$ and $b$ aren't integers, to extend the definition of $L_n^k(x)$ to non-integral $k$.  We define
\be 
L_n(x)\equiv 0 \; \text{and} \; L_n^k(x) \equiv 0 \; \text{for} \; n<0 .\ee
The associated Laguerre polynomials have the generating function
\be 
\frac{1}{(1-z)^{1+k}} \exp{\left [ -\frac{x z}{1-z}\right ]} = \sum_{n=0}^\infty z^n L_n^k(x), \quad \vert z \vert <1 , 
\ee
Rodrigues' formula
\be \label{Rodrigues} 
L_n^k(x) = \frac{e^x}{n! x^k} \frac{d^n}{dx^n} \left ( x^{n+k} e^{-x} \right ) , 
\ee
recursion relations
\bse \label{assoc_Laguerre_recursions} \bearray
xL_n^{k+1}(x) &=& (x-n) L_n^k(x) + (n+k) L_{n-1}^k(x) , \\
L_n^{k-1}(x) &=& L_n^k(x)-L_{n-1}^k(x) , \\
x L_n^{k+1}(x) &=& (n+k+1) L_{n}^k(x) - (n+1) L_{n+1}^k(x) , \\
(n+k) L_n^{k-1}(x) &=& (n+1) L_{n+1}^k(x) - (n+1-x) L_n^k(x) , \\[5pt]
\frac{d}{dx} L_n^k(x) &=& -L_{n-1}^{k+1}(x) ,
\eearray \ese
and orthogonality relation
\be 
\label{Laguerre_orthogonality} \int_0^\infty dx \, e^{-x} x^k L_n^k(x) L_m^k(x) = \frac{(n+k)!}{n!} \delta_{n m} . 
\ee
An integral formula for the associated Laguerre polynomials is
\be
L_n^k(x) = \oint \frac{dt}{2\pi i} \frac{(1+t)^{n+k}}{t^{n+1}} e^{-x t} ,
\ee
and their explicit expression in terms of hypergeometric functions is
\be
L_n^k(x) = \binom{n+k}{k} \, {_1 F_1}(-n,k+1;x) .
\ee

The Schr\"odinger-Coulomb equation in momentum space has the form
\be
\frac{\vec p\,^2 }{2m_r} \psi(\vec p\,) - \int \dbar^3 \ell \, \frac{4 \pi Z \alpha}{(\vec p - \vec \ell\,\,)^2} \psi(\vec \ell\,) = E \psi(\vec p\,)
\ee
where $\dbar^n p = \frac{d^n p}{(2 \pi)^n}$.
The momentum space Coulomb wave functions are Fourier transforms of the coordinate space functions:
\be \label{Coulomb_wf_mom} 
\psi_{n \ell m}(\vec p\,) = \int d^3 x \, e^{-i \vec p \cdot \vec x} \psi_{n \ell m}(\vec x\,) = R_{n \ell}(p) Y_{\ell m}(\hat p) 
\ee
where
\be \label{radial_fn_mom} R_{n \ell}(p) = \phi_n N_{n \ell} \frac{p^\ell \gamma_n^{\ell+1}}{D_n^{\ell+2}} C_{n-\ell-1}^{\ell+1} \!\left ( \frac{\overline{D}_n}{D_n} \right ). \ee
The constants in (\ref{radial_fn_mom}) are the coordinate space wave function at contact
\be \phi_n = \left ( \frac{\gamma_n^3}{\pi} \right )^{1/2} = \left ( \frac{(m_r Z \alpha)^3}{\pi n^3} \right )^{1/2} = \psi_{n 0 0}(\vec x = 0) \ee
and
\be N_{n \ell} = 2^{2 \ell+3} \pi \ell! \left ( \frac{4 \pi n (n-\ell-1)!}{(n+\ell)!} \right )^{1/2} .\ee
and we have used $D_n = p^2+\gamma_n^2$ and $\overline{D}_n = p^2-\gamma_n^2$.  The momentum space version of the Coulomb wave function can be obtained by Fourier transformation of the coordinate space wave function \cite{Podolsky29}, by direct solution of the momentum-space Schr\"odinger-Coulomb equation \cite{Fock36}, or from the the momentum-space Green function \cite{Schwinger64,Lieber89,Adkins92}.  Some values for the Gegenbauer polynomials are $C_0^\lambda(\beta) = 1$, $C_1^\lambda(\beta) = 2 \lambda \beta$, $C_2^\lambda(\beta) = 2 \lambda (\lambda+1) \beta^2 - \lambda$, etc.  They satisfy the Gegenbauer differential equation
\be 
\left [ (1-\beta^2) \frac{d^2}{d\beta^2} - (1+2\lambda) \beta \frac{d}{d\beta} + n (n+2\lambda) \right ] C^\lambda_n(\beta) = 0 , 
\ee
the Rodrigues formula
\be
C^\lambda_n(\beta) = \frac{(-2)^n}{n!} \frac{\Gamma(n+\lambda) \Gamma(n+2\lambda)}{\Gamma(\lambda) \Gamma(2n+2\lambda)}
(1-\beta^2)^{-\lambda+1/2} \frac{d^n}{d \beta^n} \left [ (1-\beta^2)^{n+\lambda-1/2} \right ] ,
\ee
and have the series expansion
\be
C^\lambda_n(\beta) = \sum_{k=0}^{\lfloor n/2 \rfloor} \frac{(-1)^k \Gamma(n+\lambda-k)}{\Gamma(\lambda) k! (n-2k)!} \left ( 2 \beta \right )^{n-2k} ,
\ee
where the floor $\lfloor n/2 \rfloor$ is the greatest integer in $n/2$.  The Gegenbauer generating function is
\be 
\frac{1}{(1-2 \beta x + x^2)^\lambda} = \sum_{m=0}^\infty x^m C_m^\lambda(\beta) .
\ee
The natural interval is $-1 \le \beta \le 1$, corresponding to $0 \le p < \infty$.  The Gegenbauer polynomials satisfy the orthonormality condition
\be
\int_{-1}^1 d \beta \, \left ( 1-\beta^2 \right )^{\lambda-1/2} C_n^\lambda(\beta) C_m^\lambda(\beta) 
= \frac{\pi 2^{1-2\lambda} \Gamma(n+2\lambda)}{n! (n+\lambda) \left [ \Gamma(\lambda) \right ]^2} \delta_{n m} .
\ee
Various derivative identities exist, including
\be
\frac{d}{d \beta} C_n^\lambda(\beta) = 2 \lambda C_{n-1}^{\lambda+1}(\beta) .
\ee

\section{Special functions: gamma, polygamma, beta, zeta, hypergeometric, and harmonic and diharmonic numbers}
\label{special_functions}

In this Appendix we give definitions and useful formulas for a number of special functions.  In particular, we consider the gamma, polygamma, beta, zeta, and hypergeometric functions.  We also discuss properties of the harmonic numbers that arise, for example, when expanding the gamma function near a non-positive integer.  We define ``diharmonic numbers'' as the natural generalization of harmonic numbers for use in related circumstances.  Our definitions follow the standards set, for example, in \cite{Abramowitz72,Gradshteyn80,Olver10}.

The gamma function is defined by
\be \Gamma(t) = \int_0^\infty dx \, x^{t-1} e^{-x} , \ee
so that
\be \int_0^\infty dx \, x^n e^{-\lambda x} = \frac{\Gamma(n+1)}{\lambda^{n+1}} . \ee
The gamma function satisfies $\Gamma(1)=1$ and the recursive formula
\be \Gamma(n)=(n-1) \Gamma(n-1) \ee
so that $\Gamma(n) = (n-1)!$ if $n \ge 1$ is an integer.  It also satisfies the reflection identity
\be \label{reflection} \Gamma(z) \Gamma(1-z) = \frac{\pi}{\sin(\pi z)} \ee
and the duplication formula
\be \label{duplication} \Gamma(z) \Gamma(z+1/2) = 2^{1-2z} \sqrt{\pi} \, \Gamma(2z) . \ee
Some useful values for integral and half-integral arguments are
\bse \bearray
\Gamma(1) &=& 1 , \\
\Gamma(n) &=& (n-1)! , \\
\Gamma \left ( \frac{1}{2} \right ) &=& \sqrt{\pi} , \\
\Gamma \left ( n+\frac{1}{2} \right ) &=& \frac{(2n-1)!! \sqrt{\pi}}{2^n} , \\
\Gamma \left ( -\frac{1}{2} \right ) &=& -2 \sqrt{\pi} , \\
\Gamma \left ( -n+\frac{1}{2} \right ) &=& \frac{(-2)^n \sqrt{\pi}}{(2n-1)!!} .
\eearray \ese

Some identities that prove useful in the evaluation of $\langle r^p \rangle$ for various values of $p$ and elsewhere are
\bse \bearray 
\label{gamma_fraction_identity_a} \frac{\Gamma(a+n)}{\Gamma(a)} &=& \prod_{k=0}^{n-1} (a+k) , \\
\label{gamma_fraction_identity_a2} \frac{\Gamma(-a+n)}{\Gamma(-a)} &=& (-1)^n \frac{\Gamma(a+1)}{\Gamma(a-n+1)} , \\
\label{gamma_fraction_identity_b} \frac{\Gamma(a)}{\Gamma(a-n)} &=& \prod_{k=1}^n (a-k) , \\
\label{gamma_fraction_identity_c} \frac{\Gamma(\epsilon)}{\Gamma(\epsilon-n)} &=& (-1)^n n! \left \{ 1 - \epsilon \text{H}_n + O(\epsilon^2) \right \}
\eearray \ese
where $n$ is a positive integer and $\text{H}_n$ is the harmonic number
\be \text{H}_n \equiv \sum_{k=1}^n \frac{1}{k} . \ee
We also define the generalized harmonic number $\text{H}_n^{(\alpha)}$ as
\be
\text{H}_n^{(\alpha)} \equiv \sum_{k=1}^n \frac{1}{k^\alpha} ,
\ee
which has the form of a truncated version of the sum used to define the Riemann zeta function
\be
\zeta(s) = \sum_{k=1}^\infty \frac{1}{k^s} .
\ee
Expansion of the gamma function near integral arguments can be facilitated by the formula
\be 
e^{\epsilon \gamma_E} \Gamma(1+\epsilon) = \exp \left \{ \sum_{n=2}^\infty \frac{(-1)^n \zeta(n)}{n} \epsilon^n \right \} .
\ee

The beta function $B(a,b)$ is defined as
\be B(a,b) = \frac{\Gamma(a) \Gamma(b)}{\Gamma(a+b)} . \ee
Many useful integrals can be done in terms of the beta function, including
\bse \bea
\int_0^1 dx \, x^n (1-x)^m &=& B(n+1,m+1) , \\
\int_0^{\pi/2} d \theta \, \sin^n \theta \cos^m \theta &=& \frac{1}{2} B \left ( \frac{n+1}{2}, \frac{m+1}{2} \right ) , \\
\int_0^\infty dx \, \frac{x^a}{(A+x^b)^c} &=& \frac{1}{b} A^{\frac{a+1}{b}-c} B \left ( \frac{a+1}{b}, c-\frac{a+1}{b} \right ) , \\
\int_0^\infty dx \, \frac{x^a}{(A+x^b) (B+x^b)} &=& \frac{-1}{B-A} \left ( B^{\frac{a+1}{b}-1} - A^{\frac{a+1}{b}-1} \right ) \frac{1}{b} B \left ( \frac{a+1}{b}, 1-\frac{a+1}{b} \right ) .
\eea \ese

Derivatives of the gamma function are given in terms of the polygamma functions.  Specifically, the digamma function $\psi(z)$ is the logarithmic derivative of $\Gamma(z)$:
\be 
\psi(z) \equiv \frac{\Gamma'(z)}{\Gamma(z)} 
\ee
As a direct consequence of $\Gamma(z+1)=z \Gamma(z)$ one has
\be
\psi(z+1) = \psi(z) + \frac{1}{z},
\ee
and the derived formulas
\bse \bea
\label{digamma_rising_identity} \psi(z+N) &=& \psi(z) + \sum_{r=0}^{N-1} \frac{1}{z+r} ,  \\[3pt]
\label{digamma_falling_identity} \psi(z) &=& \psi(z-N) + \sum_{r=1}^N \frac{1}{z-r} .
\eea \ese
Some useful values for integral and half-integral arguments are
\bse \bearray
\psi(1) &=& -\gamma_E , \\
\psi(n) &=& - \gamma_E + \text{H}_{n-1} , \\
\psi \left (\frac{1}{2} \right ) &=& - \gamma_E-2 \ln 2 , \\
\psi \left (n+\frac{1}{2} \right) &=& - \gamma_E - 2 \ln 2 + \sum_{k=1}^n \frac{2}{2k-1} , 
\eearray \ese
where the Euler, or Euler-Mascheroni, constant is $\gamma_E \approx 0.57721\, 56649$.
Near the origin $\psi(z)$ has a pole and the expansion
\be \psi(z) = -\frac{1}{z} - \gamma_E + \zeta(2) z + O(z^2) , 
\ee
and also, for $N$ a non-negative integer,
\bse \bea
 \label{psi_identity_1}    \lim_{\epsilon \rightarrow 0} \Bigl ( \psi(\epsilon ) - \psi(\epsilon-N) \Bigr ) &=& - \text{H}_N , \\
 \label{psi_identity_2}    \lim_{\epsilon \rightarrow 0} \frac{\psi(\epsilon-N)}{\Gamma(\epsilon-N)} &=& \lim_{\epsilon \rightarrow 0} \frac{\psi(\epsilon)}{\Gamma(\epsilon-N)}  = (-1)^{N+1} N! .
\eea \ese
The higher polygamma functions are
\be \label{eqn_def_psi_n_z}
\psi(n,z) \equiv \left ( \frac{d}{dz} \right )^n \psi(z) . 
\ee
Some useful values and relations involving the polygamma function are
\bse \bearray
\label{psi_identity_3} \psi(n,z) &=& (-1)^{n+1} n! \sum_{k=0}^\infty \frac{1}{(z+k)^{n+1}} \equiv (-1)^{n+1} n! \zeta(n+1,z) , \\
\label{psi_identity_4} \psi(n,z+1) &=& \psi(n,z) + \frac{(-1)^n n!}{z^{n+1}} , \\
\label{psi_identity_5} \psi(n,z+N) &=& \psi(n,z) + (-1)^n n! \sum_{r=0}^{N-1} \frac{1}{(z+r)^{n+1}} , \\
\label{psi_identity_6} \psi(n,z) &=& \psi(n,z-N) + (-1)^n n! \sum_{r=1}^{N} \frac{1}{(z-r)^{n+1}} .
\eearray \ese
The functions $\psi(n,z)$ have a pole of order $n+1$ near the origin and the expansion
\be
\psi(n,z) = (-1)^{n+1} n! \left \{ \frac{1}{z^{n+1}} + \zeta(n+1) + O(z) \right \} 
\ee
and satisfy identities of the sort
\be \label{psi_identity_8}
 \lim_{\epsilon \rightarrow 0} \frac{\psi(1,\epsilon-N)-\psi^2(\epsilon-N)}{\Gamma(\epsilon-N)} = \lim_{\epsilon \rightarrow 0} \frac{\psi(1,\epsilon)-(\psi(\epsilon)+\text{H}_N)^2}{\Gamma(\epsilon-N)}  = 2 (-1)^N N! (\text{H}_N-\gamma_E) .
\ee
The Hurwitz generalized zeta function $\zeta(n,x)$ is defined as
\be
\zeta(s,q) \equiv \sum_{k=0}^\infty \frac{1}{(q+k)^s} ,
\ee
and satisfies
\be
\zeta(s,N) = \sum_{k=0}^\infty \frac{1}{(N+k)^s} = \sum_{k=N}^\infty \frac{1}{k^s} = \zeta(s) - \text{H}_{N-1}^{(s)} , \quad \text{$N$ an integer with $N \ge 1$}.
\ee
with $\text{H}_0^{(s)} \equiv 0$.  The polygamma function satisfies a corresponding identity
\be
\psi(n,N) = (-1)^{(n+1)} n! \left \{ \zeta(n+1) - \text{H}_{N-1}^{(n+1)} \right \} , \quad \text{$n$, $N$ integers: $n \ge 1$, $N \ge 1$}.
\ee
In particular, one has
\bse \bea
\psi(1,1) &=& \zeta(2) , \\[3pt]
\psi(1,N) &=& \zeta(2) - \text{H}_{N-1}^{(2)} .
\eea \ese

Some useful harmonic number sums are
\bse \label{harmonic_number_sums} \bea
\sum_{r=1}^n \frac{\text{H}_r}{r} &=& \frac{1}{2} \left ( \text{H}_n^2 + \text{H}_n^{(2)} \right ) , \\[3pt]
\sum_{r=1}^{n-1} \frac{\text{H}_r}{n-r} &=& \sum_{r=1}^{n-1} \frac{\text{H}_{n-r}}{r} = \text{H}_n^2 - \text{H}_n^{(2)} ,
\eea \ese
where the latter sum comes from equation (24) of Spie{\ss} \cite{Spiess90}.

\begin{figure}
\includegraphics[width=7.0in]{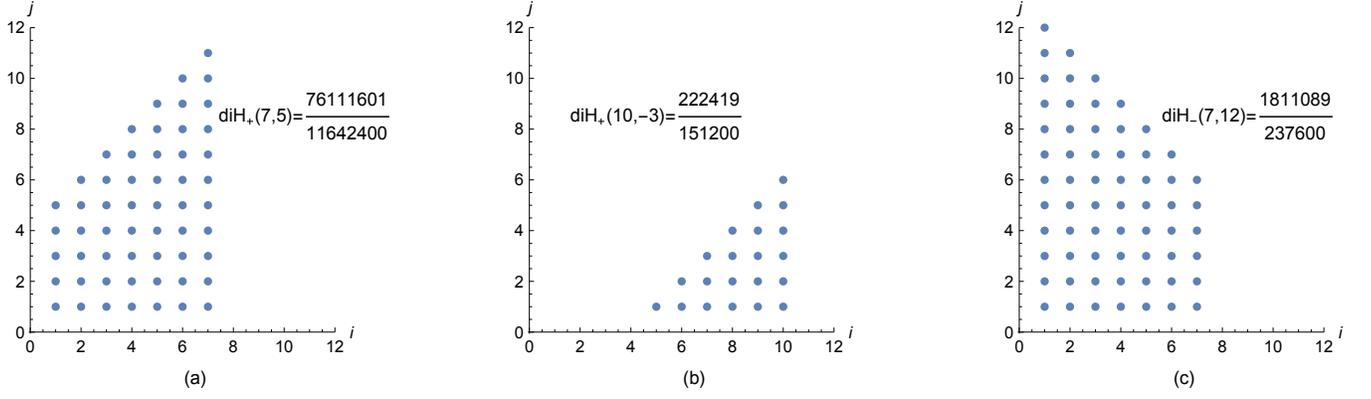}
\caption{\label{FigdiHdots} Regions in the plane of integral-valued $(i,j)$ pairs over which the sums $\sum_{\text{region}} 1/(i j)$ are defined as the diharmonic numbers: $\text{diH}_+(n,m)$ with a positively sloped diagonal segment and $\text{diH}_-(n,m)$ with a negatively sloped segment.  Part (a) shows the trapezoidal region for $\text{diH}_+(7,5)$ while (b) shows the region for $\text{diH}_+(10,-3)$, of which only the triangular tip is active in the sum.  Part (c) shows the region for $\text{diH}_-(7,12)$.  Values for the diharmonic numbers of (a)-(c) are also shown.}
\end{figure}

The ``diharmonic" sums shown above are of the form $\sum_\text{region} 1/(i j)$ where ``region'' is some region in the plane of integer pairs $(i,j)$ with $1 \le i,j < \infty$.  Diharmonic sums for rectangular regions are easy to work out since the sums factorize.  For the rectangle $a \le i \le b$, $c \le j \le d$, the diharmonic  sum is $\sum_{a \le i \le b, \; c \le j \le d} 1/(i j) = (\text{H}_b-\text{H}_{a-1})(\text{H}_d-\text{H}_{c-1})$.  More generally, in analogy to the dilogarithm function, we define the rising and falling ``diharmonic numbers''
\bse \label{diharmonic_defs} \bea
\text{diH}_+(n,m) &\equiv& \sum_{i=1}^n \frac{\text{H}_{m-1+i}}{i} = \sum_{i=1}^n \sum_{j=1}^{m-1+i} \frac{1}{i j} \quad (\text{non-zero for } n \ge 1, m \ge 2-n) , \\[3pt]
\text{diH}_-(n,m) &\equiv& \sum_{i=1}^n \frac{\text{H}_{m+1-i}}{i} = \sum_{i=1}^n \sum_{j=1}^{m+1-i} \frac{1}{i j} \quad (\text{non-zero for } n \ge 1, m \ge 1) . 
\eea \ese
The regions in the $(i,j)$ plane over which the sums of $1/(i j)$ are taken are illustrated in Fig.~\ref{FigdiHdots}.  For the purposes of these definitions, we assume that $\text{H}_n=0$ for $n \le 0$ and that the summations vanish when the upper limit is less than the lower.  The diharmonic sums allow us to find the values of the sum of $1/(i j)$ over any polygonal region of $(i,j)$--space with $1 \le i,j < \infty$ having sides that are either vertical, horizontal, or diagonal at $45$ degrees.  Some special cases come from (\ref{harmonic_number_sums}) above
\be
\text{diH}_+(n,1) = \sum_{i=1}^n \frac{\text{H}_i}{i} = \frac{1}{2} \left ( \text{H}_n^2 + \text{H}_n^{(2)} \right ) 
\ee
and
\be
\text{diH}_-(n-1,n-1) = \sum_{i=1}^{n-1} \frac{\text{H}_{n-i}}{i} = \text{H}_n^2 - \text{H}_n^{(2)} .
\ee
The diharmonic numbers are much more general, as the special cases both come from regions with a vertex at $i=j=1$ and with one side along the horizontal axis.  Some more general sums appear in Spie{\ss}, but apparently only for triangular regions with the diagonal on or close to the line $i=j$.  The diharmonic sums for arbitrary polygonal regions in the $(i,j)$ plane having all sides either horizontal, vertical, or diagonal at $45$ degrees can be obtained as linear combinations of $\text{diH}_\pm(n,m)$ and the sums for rectangular regions.  Closed forms for diharmonic numbers of nearby regions can be obtained using the recursion relations
\bse \bea
\text{diH}_+(n+1,m) &=& \text{diH}_+(n,m) + \frac{1}{n+1} \text{H}_{n+m} , \\[3pt]
\text{diH}_+(n,m+1) &=& \begin{cases} \text{diH}_+(n,m) + \frac{1}{m} \left ( \text{H}_n + \text{H}_m - \text{H}_{n+m} \right ) & \text{if  } m \ge 1 , \\[3pt]
\text{diH}_+(n,0) + \text{H}_n^{(2)} & \text{if } m=0 , \\[3pt]
\text{diH}_+(n,m) + \frac{1}{m} \left ( \text{H}_n - \text{H}_{-m} - \text{H}_{n+m} \right ) & \text{if  } -n \le m \le -1 , \end{cases}  \\[3pt]
\text{diH}_-(n+1,m) &=& \begin{cases} \text{diH}_-(n,m) & \text{if } n \ge m , \\[3pt] \text{diH}_-(n,m) + \frac{1}{n+1} \text{H}_{m-n} & \text{if } n < m , \end{cases} \\[3pt]
\text{diH}_-(n,m+1) &=& \begin{cases} \text{diH}_-(n,m) + \frac{1}{m+2} \left ( \text{H}_{m+1}+\text{H}_n-\text{H}_{m+1-n} \right ) & \text{if  } m \ge n-1 , \\[3pt]
\text{diH}_-(n,m) + \frac{2}{m+2} \text{H}_{m+1} & \text{if  } m < n-1 . \end{cases}
\eea \ese
Values for diharmonic numbers of nearly-triangular regions include:
\bse \bea
\text{diH}_+(n,2) &=& \half \left ( \text{H}_{n}^2 + \text{H}_n^{(2)} \right ) + 1 - \frac{1}{n+1} , \\[3pt]
\text{diH}_+(n,1) &=& \half \left ( \text{H}_{n}^2 + \text{H}_n^{(2)} \right ) , \\[3pt]
\text{diH}_+(n,0) &=& \half \left ( \text{H}_n^2 - \text{H}_n^{(2)} \right ) , \\[3pt]
\text{diH}_+(n,-1) &=&  \half \left ( \text{H}_n^2 - \text{H}_n^{(2)} \right ) - 1 + \frac{1}{n} , \\[3pt]
\text{diH}_-(n-1,n-1) &=& \text{H}_n^2 - \text{H}_n^{(2)} , \\[3pt]
\text{diH}_-(n-1,n) &=& \text{H}_{n+1}^2 - \text{H}_{n+1}^{(2)} - \frac{1}{n} , \\[3pt]
\text{diH}_-(n-1,n+1) &=&\text{H}_{n+2}^2 - \text{H}_{n+2}^{(2)} - \frac{1}{n} - \frac{1}{n+2} \left ( \frac{1}{n} + \frac{1}{n+1} + \frac{3}{2} \right ) .
\eea \ese
The diharmonic numbers also satisfy reflection identities
\bse \bea
\text{diH}_+(n,m) &=& \text{H}_{n+m-1} \text{H}_n - \text{diH}_+(n+m-1,-m+1) , \\[3pt]
\text{diH}_-(n,m) &=& \text{H}_{m+1}^2 - \text{H}_{m+1}^{(2)} - \text{diH}_-(m-n+1,m) + \text{H}_{m-n+1} \text{H}_n .
\eea \ese
found by consideration of the regions in $(i,j)$ space covered by the corresponding sums.

The hypergeometric function $_2 F_1(a,b; c; x)$ is defined by
\be
_2 F_1(a,b;c;x) = \sum_{n=0}^\infty \frac{a^{\bar n} b^{\bar n}}{c^{\bar n}} \frac{x^n}{n!} ,
\ee
and more generally by
\be
{_p}F_q(a_1, \dots , a_p; c_1 , \dots , c_q ; x) = \sum_{n=0}^\infty \frac{a_1^{\bar n} \cdots a_p^{\bar n}}{c_1^{\bar n} \cdots c_q^{\bar n}} \frac{x^n}{n!} ,
\ee
where the rising and falling factorials are
\bse \bea
a^{\bar n} &\equiv& a (a+1) \cdots (a+n-1) = \frac{\Gamma(a+n)}{\Gamma(a)} = n! \binom{a+n-1}{n} \quad \text{(rising factorial)} , \\[3pt]
a^{\underline n} &\equiv& a (a-1) \cdots (a-n+1) = \frac{\Gamma(a+1)}{\Gamma(a-n+1)} = n! \binom{a}{n} \quad \text{(falling factorial)} .
\eea \ese
(The rising factorial is also known as the ``Pochhammer symbol'': $a^{\bar n} = (a)_n$.  Other notations for the rising and falling factorials are $a^{\bar n} = a^{(n)}$ and, confusingly, $a^{\underline n} = (a)_n$.  One must always verify definitions when dealing with rising and falling factorials.  The advantage of the notation chosen here is that it is unambiguous and suggestive of its meaning.)  Rising and falling factorials satisfy identities reminiscent of the binomial formula:
\bse \bea
(a+b)^{\bar n} &=& \sum_{r=0}^n \binom{n}{r} a^{\overline{n-r}} \, b^{\overline r} , \\[3pt]
(a+b)^{\underline{n}} &=& \sum_{r=0}^n \binom{n}{r} a^{\underline{n-r}} \, b^{\underline{r}} .
\eea \ese
The hypergeometric series terminates when $a$ or $b$ is a negative integer.  A hypergeometric function identity that is useful when doing sums involving factorials and binomial coefficients is given as (5.93) of Graham, Knuth, and Patashnik \cite{Graham94}:
\be \label{hypergeometric_identity_1}
_2 F_1(a,-n;c;1) = \frac{\Gamma(c-a+n) \Gamma(c)}{\Gamma(c-a) \Gamma(c+n)} = \frac{(c-a)^{\bar n}}{c^{\bar n}} = \frac{(a-c)^{\underline{n}}}{(-c)^{\underline{n}}}  \quad \text{for integer $n \ge 0$}.
\ee
One often requires series expansions of $_2 F_1(a,b;c;x)$ when $a$, $b$, and $c$ differ from integers by terms of order $\epsilon$ with one of those integers in $a$ or $b$ being negative.  For example, when $a \rightarrow -n+a \epsilon$, $b \rightarrow k+b \epsilon$, $c \rightarrow -n+1+a \epsilon$, and $x \rightarrow -1$, with $n$ an integer $n \ge 1$ and $k$ an integer $k \ge 0$, we define
\be
f(n,k,a,b) \equiv {_2 F_1}(-n+a \epsilon, k+b \epsilon;-n+1+a \epsilon; -1) .
\ee
For the case $n \ge 1$, $k \ge 1$, one has
\bea
f(n \ge 2,k \ge 1,a,b) &=& \sum_{j=0}^\infty \frac{(-1)^j (-n+a \epsilon)^{\bar j} (k+b \epsilon)^{\bar j}}{(-n+1+a \epsilon)^{\bar j} j! } \crr
&=& 1 + \sum_{j=1}^\infty \frac{(-1)^j (-n+a \epsilon) (k+b \epsilon)^{\bar j}}{(-n+j+a \epsilon) j! } \crr
&=& 1 + (-n+a \epsilon) \left \{ \sum_{j=1}^{n-1} \frac{(-1)^j k^{\bar j}}{(-n+j) j! } + \frac{(-1)^n (k+b \epsilon)^{\bar n}}{(a \epsilon) n!} + \sum_{j=n+1}^\infty \frac{(-1)^j k^{\bar j}}{(-n+j) j! } + O(\epsilon) \right \} ,
\eea
where sums are taken to be zero when the upper limit is smaller than the lower limit.  The first sum here is easy to evaluate for given $n$ and $k$.  The second term can be expanded in $\epsilon$ using
\bea
\frac{(k+b \epsilon)^{\bar n}}{\epsilon} &=& \frac{1}{\epsilon} \left \{ k \left (1+\frac{b \epsilon}{k} \right ) (k+1) \left ( 1 + \frac{b \epsilon}{k+1} \right ) \cdots (k+n-1) \left ( 1 + \frac{ b \epsilon}{k+n-1} \right ) \right \} \crr
&=& \frac{(k+n-1)!}{(k-1)!} \left \{ \frac{1}{\epsilon} + b \left ( \text{H}_{k+n-1} - \text{H}_{k-1} \right ) + O(\epsilon) \right \} .
\eea
The final infinite sum evaluates to
\be
\sum_{j=n+1}^\infty \frac{(-1)^j k^{\bar j}}{(-n+j) j! } = \frac{(-1)^{n+1} k^{\overline{n+1}}}{(n+1)!} {_3 F_2}(1,1,n+k+1; 2,n+2; -1) ,
\ee
which, at least for integers $n$, $k$ with $0 \le n,k \le 16$, evaluates to a form $\alpha \ln 2 + \beta$ for some $\alpha$, $\beta$.  So in all one has
\bea \label{hyperf_case1}
f(n \ge 1,k \ge 1,a,b) &=& \frac{(-1)^{n+1} n}{a \epsilon} \binom{k+n-1}{n} + 1 + (-1)^n \binom{k+n-1}{n} \left [ 1-\frac{n b}{a} \left ( \text{H}_{k+n-1}-\text{H}_{k-1} \right ) \right ] \crr
&\hbox{}& \quad + n \sum_{j=1}^{n-1} \frac{(-1)^j k^{\bar j}}{(n-j) j! } + \frac{(-1)^{n} n k^{\overline{n+1}}}{(n+1)!} {_3 F_2}(1,1,n+k+1; 2,n+2; -1) + O(\epsilon) .
\eea
Other cases require separate analyses, which proceed along similar lines.  For instance, with $n \ge 1$, $k=0$ one finds
\be \label{hyperf_case2}
f(n \ge 1,0,a,b) = \left ( 1-(-1)^n \frac{b}{a} \right ) + b \epsilon \left \{ \left (-1+(-1)^n \right ) \ln 2 - (-1)^n \frac{b}{a} \text{H}_{n-1} + \sum_{j=1}^{n-1} \frac{(-1)^j}{n-j} \right \} + O(\epsilon^2) .
\ee

\section{Evaluation of the bracket $\blangle \! \ln q \brangle_{\vec p_2,\vec p_1}$}
\label{eval_of_bracket_lnq}

The momentum space bracket $\blangle \! \ln q \brangle_{\vec p_2,\vec p_1}$ is easy to evaluate for any specific state via momentum space integration.  This is a finite bracket with no need for regularization, so we calculate in three dimensions.  In order to obtain a general formula for this bracket, it is best to work with coordinate space expectation values.  However, the three dimensional Fourier transform of $\ln q$ is $\text{FT}[\ln q](\vec x\,) = -1/(4 \pi r^3)$, which has a divergent expectation value for S states.  Dimensional regularization fails to regulate the Fourier transform of $\ln q$, which is proportional to $1/r^D$ in $D$ dimensions, so we stick to three dimensions and regulate differently.  We note that $\frac{d}{ds} q^s = q^s \ln q $ and
\be
\ln q = \lim_{s \rightarrow 0} \frac{d}{ds} q^s ,
\ee
so we use $s$ as a regulating parameter and hold off on the derivative and $s \rightarrow 0$ limit until after taking the expectation value.  That is, we evaluate $\blangle \!\ln q \brangle_{\vec p_2,\vec p_1}$ as
\be
\blangle \! \ln q \brangle_{\vec p_2,\vec p_1} = \lim_{s \rightarrow 0} \frac{d}{ds} \blangle \text{FT}[q^s](\vec x\,) \brangle .
\ee
The three-dimensional Fourier transform of $q^s$, from Appendix~\ref{Fourier_transform}, is
\be
\text{FT}[q^s](\vec x\,) = \int \dbar^3 q \, e^{i \vec q \cdot \vec x} q^s = \frac{2^s \Gamma \left ( \frac{3+s}{2} \right )}{\pi^{3/2} \Gamma \left (-\frac{s}{2} \right ) r^{3+s}} ,
\ee
with the expectation value
\bea
\blangle \text{FT}[q^s](\vec x\,) \brangle_{n 0} &=& \int dr \, r^2 R_{n 0}(r) \text{FT}[q^s](\vec x\,) R_{n 0}(r) \crr
&=& \frac{2^s \Gamma \left ( \frac{3+s}{2} \right )}{\pi^{3/2} \Gamma \left (-\frac{s}{2} \right )} \left ( \frac{2}{a n} \right )^s \left ( \frac{4}{a^3 n^5} \right ) \int_0^\infty d\rho \, e^{-\rho} \rho^{-s-1} \left ( L_{n-1}^1(r) \right )^2 \crr
&=& \left ( \frac{4}{a n} \right )^s \frac{\Gamma \left ( \frac{3+s}{2} \right )}{\pi^{3/2} \Gamma \left (-\frac{s}{2} \right )} \left ( \frac{4}{a^3 n^5} \right ) K_{-s-1}(n-1,1;n-1,1) \crr
&=& \left ( \frac{4}{a n} \right )^s \frac{\Gamma \left ( \frac{3+s}{2} \right )}{\pi^{3/2} \Gamma \left (-\frac{s}{2} \right )}  \left ( \frac{4}{a^3 n^5} \right ) \frac{n! (-1)^{n-1}}{(n-1)!} \sum_{r=0}^{n-1} \frac{(-1)^r \Gamma(-s-1+r) \Gamma(-s+r)}{r! (n-1-r)! (r+1)! \Gamma(-s+r-n)} ,
\eea
after use of the S-state wave function of (\ref{Sradial_wf}) and the integration formula (\ref{gen_Laguerre_int}).  Application of an $s$ derivative and the limit $s \rightarrow 0$ yields the S-state bracket of $\ln q$:
\bea
\blangle \! \ln q \brangle_{\vec p_2,\vec p_1} &=& \lim_{s \rightarrow 0} \left ( \frac{4}{a n} \right )^s \frac{\Gamma \left ( \frac{3+s}{2} \right )}{\pi^{3/2} \Gamma \left (-\frac{s}{2} \right )}  \left ( \frac{4}{a^3 n^5} \right ) \frac{n! (-1)^{n-1}}{(n-1)!} \sum_{r=0}^{n-1} \frac{(-1)^r \Gamma(-s-1+r) \Gamma(-s+r)}{r! (n-1-r)! (r+1)! \Gamma(-s+r-n)} \crr
&\hbox{}& \hspace{-0.4cm} \times \left \{ \ln \left ( \frac{4}{a n} \right ) + \frac{1}{2} \psi \left ( \frac{3+s}{2} \right ) - \psi(-s-1+r) - \psi(-s+r) + \frac{1}{2} \psi \left ( - \frac{s}{2} \right ) + \psi(-s+r-n) \right \} .
\eea
We can simplify the expression in curly brackets slightly by use of $\psi(-s-1+r)-\psi(-s+r-n) = \sum_{j=2}^n \frac{1}{r-s-j}$, which has a linear divergence for $s \rightarrow 0$ when $j=r$.  In the $s \rightarrow 0$ limit $\Gamma(-s/2)$ has a linear divergence, as does $\Gamma(-s-1+r)$ for $r=0$ and $r=1$, $\Gamma(-s+r)$ for $r=0$, and $\Gamma(-s+r-n)$ for $0 \le r \le n-1$.  Near $s=0$, the expression in curly brackets is $\ln \left (\frac{2}{a n} \right ) + H_n + O(s)$ for $r=0$, $1/s + O(s^0)$ for r=1, and $O(1/s)$ for $2 \le r \le n-1$.  We see that the only terms in the sum over $r$ that contribute to the $s \rightarrow 0$ limit have $r=0$ and $r=1$, and we find
\be
\blangle \! \ln q \brangle_{\vec p_2,\vec p_1} = \frac{(m_r Z \alpha)^3}{\pi n^3} \left \{ \ln \left ( \frac{2 m_r Z \alpha}{n} \right ) + H_n + \frac{n-1}{2n} \right \}
\ee
for S states.  We have checked this formula for $\blangle \! \ln q \brangle_{\vec p_2,\vec p_1}$ by explicit integration in momentum space for $1 \le n \le 36$.  For $\ell > 0$ the bracket $\blangle \! \ln q \brangle_{\vec p_2,\vec p_1} = - \frac{1}{4\pi} \blangle r^{-3} \brangle$ is finite and can be evaluated in the usual way.  Our result for $\blangle \! \ln q \brangle_{\vec p_2,\vec p_1}$ agrees with that of Titard and Yndur\'ain \cite{Titard94}.


\begin{acknowledgments}
We acknowledge the support of the National Science Foundation through Grant No. PHY-1707489 and of the Franklin \& Marshall College Hackman Scholars Program.
\end{acknowledgments}
 


\end{document}